\documentclass[pdflatex,sn-mathphys-num]{sn-jnl}
\usepackage{epsfig}
\usepackage{mathrsfs}
\usepackage{array,longtable}
\newcolumntype{C}{>{$\displaystyle}l<{$}} 

\usepackage{amsfonts}
\usepackage{color}
\usepackage{amssymb}

\usepackage{siunitx}
\usepackage{amsfonts}
\usepackage{amsmath}
\usepackage{nicefrac}
\usepackage{physics} 
\usepackage{slashed}
\usepackage{framed}
\usepackage{mathtools}
\usepackage{dirtytalk}
\usepackage[dvipsnames]{xcolor}
\usepackage[caption=false]{subfig}

\usepackage{hyperref}
\usepackage{xstring}

\usepackage{ulem}

\newcommand{\IS}{\mspace{2mu}} 

\allowdisplaybreaks

\usepackage{graphicx}
\usepackage{subdepth} 
\usepackage{booktabs}

\begin{document}
\newcommand{\renQ}[4][Q]{\ensuremath{\bar{I}_{#1}^{(#2,#3,#4)}}} 

\newcommand{\renQR}[4][QR]{\ensuremath{\bar{I}_{#1}^{(#2,#3,#4)}}} 
\newcommand{\renLQ}[4][LQ]{\ensuremath{\bar{I}_{#1}^{(#2,#3,#4)}}}
\newcommand{\renLQR}[6][L-Q-R]{\ensuremath{\bar{I}_{\StrBefore{#1}{-}^{#2}\StrBetween[1,2]{#1}{-}{-}\StrBehind[2]{#1}{-}^{#3}}^{(#4,#5,#6)}}}
\newcommand{\renSQR}[2][QR]{\ensuremath{\bar{S}_{#1}^{(#2)}}} 
\newcommand{\renSLQ}[2][LQ]{\ensuremath{\bar{S}_{#1}^{(#2)}}}
\newcommand{\renSLQR}[4][L-Q-R]{\ensuremath{\bar{S}_{\StrBefore{#1}{-}^{#2}\StrBetween[1,2]{#1}{-}{-}\StrBehind[2]{#1}{-}^{#3}}^{(#4)}}}

\newcommand{\alpA}[3]{\ensuremath{\alpha^{A,#3}_{#1,#2}}} 

\newcommand{\kernelQ}[3][Q]{\ensuremath{K_{#1,#2}^{(#3)}}}\
\newcommand{\kernelR}[3][R]{\ensuremath{K_{#1,#2}^{(#3)}}}
\newcommand{\kernelL}[3][L]{\ensuremath{K_{#1,#2}^{(#3)}}}

\newcommand{\kernelQR}[4][Q-R]{\ensuremath{K_{\StrBefore{#1}{-}\StrBehind{#1}{-}^{#2},#3}^{(#4)}}}
\newcommand{\kernelLQ}[4][L-Q]{\ensuremath{K_{\StrBefore{#1}{-}^{#2}\StrBehind{#1}{-},#3}^{(#4)}}}
\newcommand{\kernelSigma}[5][LR]{\ensuremath{\Sigma_{(#1)^{(#2,\,#3)},#4}^{(#5)}}}
\newcommand{\kernelLQR}[5][L-Q-R]{\ensuremath{K_{\StrBefore{#1}{-}^{#2}\StrBetween[1,2]{#1}{-}{-}\StrBehind[2]{#1}{-}^{#3},#4}^{(#5)}}}

\setlength{\parindent}{0pt}	
\setlength{\parskip}{0pt}	

\title{Finite-box effects in the axial-vector form factor:  \\ a case study for the nucleon}

\author[1,2]{\fnm{Felix} \sur{Hermsen}}

\author[2,3]{\fnm{Tobias} \sur{Isken}}

\author*[2]{\fnm{Matthias F. M.} \sur{Lutz}}\email{m.lutz@gsi.de}
\author[1]{\fnm{Rob G. E.} \sur{Timmermans}}

\affil[1]{\orgdiv{Van Swinderen Institute for Particle Physics and Gravity}, \orgname{University of Groningen}, \orgaddress{\street{Nijenborgh 9}, \postcode{9747 AG} \state{Groningen}, \country{The Netherlands}}}

\affil[2]{\orgdiv{GSI Helmholtzzentrum f\"ur Schwerionenforschung GmbH}, \orgaddress{\street{Planckstra\ss e 1}, \postcode{64291} \city{Darmstadt}, \country{Germany}}}

\affil[3]{\orgdiv{Helmholtz Forschungsakademie Hessen f\"ur FAIR (HFHF)}, \orgaddress{\street{Campus Darmstadt} \postcode{64291}, \country{Germany}}}

\abstract{
We consider the axial-vector form factor of the nucleon in a finite box. Starting from the chiral Lagrangian with nucleon and $\Delta$-isobar degrees of freedom, we address, at the one-loop level, the impact of two types of finite-volume effects. On the one hand, there are the implicit effects from the in-box values of the nucleon and $\Delta$-isobar masses. On the other hand, there are the explicit effects caused by computing the in-box loop integrals with the values of the nucleon and $\Delta$-isobar masses obtained in the infinite-volume limit. Selected numerical results are shown for three lattice ensembles. We show that the implicit effects dominate the in-box form factor. Our results are presented in terms of a set of basis functions that generalize the Passarino-Veltman reduction scheme to the finite-box case, such that only scalar loop integrals have to be performed. The techniques we developed are more generally relevant for lattice studies of hadronic quantities.
}

\maketitle

\newpage
\section{Introduction}	
An accurate evaluation of the axial-vector form factor of the nucleon from QCD with a lattice approach to its path-integral representation is an ongoing challenge for the hadron physics community \cite{Beane:2004rf,Khan:2006de,Bali:2014nma,Capitani:2017qpc,Alexandrou:2017hac,Bali:2018qus}. Despite considerable progress there remain important not sufficiently well understood issues. While by now it is possible to simulate such systems at physical pion masses \cite{Alexandrou:2023qbg,Tsuji:2024scy}, excited-state contaminations \cite{Gupta:2024qip} in the extraction of the asymptotic signal from Euclidean time simulations are increasingly difficult to control when the pion mass gets smaller, approaching its physical value. If ensembles with non-physical values are used (e.g. Refs. \cite{Khan:2006de,Beane:2004rf,Bali:2014nma,Capitani:2017qpc,Alexandrou:2017hac,Bali:2018qus,Djukanovic:2024krw,Bali:2023sdi,Gupta:2018qil}), extrapolation to the physical pion mass but also from the finite-box results to the infinite volume is needed.   Often these extrapolations are done at a phenomenology level, in particular with the finite-volume effects proportional to $m_\pi^2\,\exp(-m_\pi\,L)/\sqrt{m_\pi\,L}$, see e.g. Refs. \cite{Djukanovic:2024krw,Bali:2023sdi,Bali:2022qja,Gupta:2018qil,Park:2025rxi}. As already emphasized in 2012 \cite{Greil:2011aa}, but also in recent works \cite{Lutz:2014oxa,Lutz:2018cqo,Lutz:2020dfi,Lutz:2023xpi,Hermsen:2024eth,Hall:2025ytt} a controlled extrapolation is much more demanding, especially if the $\Delta$-isobar degree of freedom plays a critical role in the physics of the hadronic observables under consideration. 

The use of the flavor-SU(2) chiral Lagrangian formulated with nucleon and $\Delta$-isobar degrees of freedom \cite{Jenkins:1990jv,Jenkins:1991es,Fuchs:2003vw,Procura:2006gq,Ledwig:2011cx,Yao:2017fym,Lutz:2020dfi,Hermsen:2024eth,Alvarado:2021ibw,Alvarado:2021ibw} is expected to help making further progress in this extrapolation challenge. From a conceptual point of view an application of the flavor-SU(2) chiral Lagrangian to Lattice-QCD (LQCD) ensembles at fixed physical strange-quark mass, but varying values of the light-quark masses, appears most promising. So far, mostly Coordinated Lattice Simulations (CLS) ensembles are available, where the sum of up-, down-, and strange-quark masses are held constant. Recently, the Mainz group performed a chiral-SU(2) extrapolation to such data, however, the dependence of the low-energy constants (LECs) on the varying strange-quark mass was not taken into account explicitly \cite{Djukanovic:2024krw}. Moreover, size and even the sign of the required chiral correction effects appears to contradict the expectation of conventional Chiral Perturbation Theory (ChPT) \cite{Djukanovic:2024krw,Harris:2019bih,Ottnad:2022axz,Capitani:2017qpc,Bali:2023sdi}.
\\
\par
The intricate interplay of a combined chiral and large-volume extrapolation was studied on somewhat older LQCD data on flavor-SU(2) ensembles  \cite{Alexandrou:2008tn,Alexandrou:2010hf,Capitani:2017qpc,Capitani:2015sba,Bali:2018qus} in Refs. \cite{Lutz:2020dfi,Hermsen:2024eth}. A successful reproduction of the axial-vector and the induced pseudoscalar form factor was achieved. 
An evaluation of both form factors from the chiral-SU(2) Lagrangian showed that the impact of the pion-$\Delta$ loop effects is significant. In particular, a large sensitivity on the choice made for the $\Delta$-isobar mass inside the loop contributions was instrumental to describe the LQCD data set. In turn, striking implicit finite-volume effects result from using the in-box hadron masses inside the one-loop contributions were the consequence. Various lattice groups \cite{Alexandrou:2023qbg,Djukanovic:2022wru} report a non-detectable finite-volume effect in the axial-vector form factor if $m_\pi\,L\geq 3.5$.  Such findings are in contradiction to our previous work \cite{Lutz:2020dfi}, where only finite-volume effects due to the in-box hadron masses were considered. This tension implicates the need of a further investigation of the full finite-volume effects originating from a one-loop calculation.  
\\
\par
A thorough study of finite-volume effects from in-box loop contributions in terms of on-shell hadron masses has not been done so far.  Of course, finite-volume corrections calculations are available to different extents \cite{Beane:2004rf,Khan:2006de,Lozano:2020qcg,Liang:2022tcj,Lutz:2014oxa,Hasenfratz:1989pk,Greil:2011aa}. In Ref. \cite{Beane:2004rf} only the axial charge $g_A$ was considered in heavy-baryon ChPT, including the $\Delta$-resonance as an explicit
degree of freedom, while in  Ref. \cite{Khan:2006de} the small-scale expansion was used to determine $g_A$. We will derive results that include the momentum dependence of the form factor. The use of on-shell hadron masses inside our loop functions departs from the standard ChPT approach, but  has significant impact on the converging properties of the chiral expansion \cite{Lutz:2018cqo,Lutz:2020dfi,Isken:2023xfo}.
\\ \par
With this work we extend our results of Refs. \cite{Lutz:2020dfi,Hermsen:2024eth} and derive a new minimal set of in-box basis integrals, which generalize our newly established chiral reduction scheme \cite{Isken:2023xfo} from the infinite volume to the finite volume. In particular, we will decompose  the one-loop integrals contributing to the axial-vector form factor of the nucleon into such basis integrals.  This paper is organized as follows: In Section \ref{sec:Axial} we discuss the framework to calculate the axial-vector form factor of the nucleon and the integrals that we need to evaluate in the finite box. Next, in Section \ref{sec:Basis} we construct the set of basis functions used for the different loop integrals. In Section \ref{sec:Numerical} we present numerical results for the contributions to the loop integrals for several lattice ensembles. We summarize in Section \ref{sec:Summary}. Two Appendices are devoted to various technical developments and definitions not given in the main text.

\section{The axial-vector form factor of the nucleon in a box}
\label{sec:Axial}
In the infinite volume, the matrix element of the axial current of QCD between on-shell nucleon states can be decomposed into two form factors as \cite{GASSER1988779,Weinberg:1958ut}
\begin{eqnarray}
 &&\bra{N(\bar{p})}A_{i}^{\mu}(0)\ket{N(p)} =
\bar{u}_{N}(\bar{p})\,F_A^\mu (q^2)\,\frac{\tau_i}{2}\,u_N(p) \,,
\nonumber \\
&&F_A^\mu(\bar{p},p) =\Big(\gamma^{\mu}\,G_A(q^2)+
\frac{q^{\mu}}{2\,M_N} \, G_P(q^2)\Big) \,\gamma_5\, ,
\label{Axialcurrent}
\end{eqnarray}
where $q=\bar{p}-p$ and $\bar p^2=p^2=M_N^2$. The functions $G_A(q^2),\, G_P(q^2)$ are the axial-vector and the induced pseudoscalar form factor of the nucleon, respectively. We assume exact isospin symmetry, with the conventional Pauli matrices $\tau_i$.\\\par

We now turn to the case of a cubic finite box with the volume $L^3$ and periodic boundary conditions. In the finite box only discrete momenta are allowed. Thus we need to replace the loop momentum integration by
\begin{equation}
    \int\frac{\dd[3]\vec{l}}{(2\,\pi)^3}\,f(\vec{l}) \quad \to \quad \frac{1}{L^3}\,\sum_{\vec{n}\in\mathbb{Z}^3}\,f(\vec{l}_n)\,,\qquad \vec{l}_n = \frac{2\,\pi}{L}\,\vec{n}\,,
    \label{eqn:ReplaceMomentaInBox}
\end{equation}
as it commonly done, see e.g. \cite{Greil:2011aa,Doring:2011ip,Lozano:2020qcg}. It is most convenient to use the Poisson summation formula to rewrite the right hand side of Eq. \eqref{eqn:ReplaceMomentaInBox}
\begin{equation}
\int\frac{\dd{l_0}}{2\,\pi}\,\frac{1}{L^3}\,\sum_{\vec{n}\in\mathbb{Z}^3}\,f(l_0,\vec{l}_n) = \sum_{\vec{n}\in\mathbb{Z}^3}\,\int\,\frac{\dd[4]{l}}{(2\,\pi)^4}\,e^{i\,\vec{l}\cdot \vec{x}_n}\,f(l_0,\vec{l})\,,
\end{equation}
where we introduced $\vec{x}_n = L\,\vec{n}$. Thus, we need to generalize the decomposition of the form factor in Eq. \eqref{Axialcurrent} to the finite box case: 
\begin{equation}\begin{aligned}[b]
&F^{\mu}_A(\bar{p},p) \xrightarrow{\text{finite box}} \Big(\gamma^\mu\,G_A(\bar{p},p) + \frac{q^\mu}{2\,M_N}\,G_P(\bar{p},p) + \frac{Q^{\mu}}{2\,M_N}\,G_T(\bar{p},p) 
\\ &\hspace{10em}
+ 2\,M_N\,\sum_{\vec{n}\in \mathbb{Z}^{3}}^{n \neq 0}\,x_n^{\mu}\, G_X(\bar{p},p,\vec{x}_n)\,\Big)\,\gamma_5\,,
\label{eqn:AxialCurrentFVE}
\end{aligned}\end{equation}
where $Q^{\mu} = \bar{p}^{\mu} + p^{\mu}$ and $x_n^{\mu} = (0,\vec{x}_n)$. In anticipation of our ongoing work \cite{Hermsen:2025} we expect this relation to hold in general. Although at the physical point the form factors depend only on $q^2$, the finite box introduces a dependence on the three-momenta $\vec {\bar p}$ and $\vec p$. The form factor $G_T(\bar{p},p)$ is sometimes referred as the induced pseudotensorial form factor and is accompanied by the structure $i\,\sigma^{\mu\nu}\,q_{\nu}$ see e.g. \cite{Schindler:2006jq,Weinberg:1958ut}. In the isospin conserved case, the Dirac structures $Q^{\mu}$ and $i\,\sigma^{\mu\nu}\,q_{\nu}$ are equivalent, so we choose $Q^{\mu}$ out of simplicity. For conserved G-parity, the form factors 
satisfy the following relations
\begin{eqnarray}
&& G_A(\bar{p},p) = +\,G_A(p,\bar{p} ) \,, \qquad
    G_P(\bar{p},p) = +\,G_P(p,\bar{p} ) \,,
    \qquad{\rm but }\quad
\nonumber\\
&&  G_T(\bar{p},p) = -\,G_T(p,\bar{p} ) \,, \qquad
    G_X(\bar{p},p, \vec x_n) = -\,G_X(p,\bar{p},\vec x_n ) \,.
\end{eqnarray}
In particular 
$G_T(\bar{p},p)$  may be non-vanishing for $p \neq \bar p$.\\\par

It is convenient to express the form factors as a projection in terms of a trace over the amplitude $F_A^\mu(\bar p, p)$ in $d$ space-time dimensions \cite{Sauerwein:2021jxb}. Specifically, we use
\begin{eqnarray}
G_A(\bar p, p) = 
&&\frac{2\,\sqrt{2} \, M_N^2}{(d-2)\,\big(t - 4\,M_N^2\big)} \, 
\Tr\bigg[\gamma_{\mu} \, \gamma_5 \, 
\frac{\slashed{\bar{p}}+M_N}{2 \, M_N} \, F_A^{\mu}(\bar p,p) \, \frac{\slashed{p}+M_N}{2 \, M_N}\bigg]
\nonumber\\ 
&&+\,\frac{8\,\sqrt{2}\,M_N^4}
{(d-2)\,\big(t - 4\,M_N^2\big)\,t}
\Tr\bigg[\frac{q_\mu}{2\,M_N} \, \gamma_5\;
\frac{\slashed{\bar{p}}+M_N}{2 \, M_N} \,  F_A^{\mu}(\bar p,p) \, \frac{\slashed{p}+M_N}{2 \, M_N}\bigg] \,,
\label{ProjAxialcurrent}
\end{eqnarray}
which holds for both the infinite-volume but also for the finite box case. How the projectors for the remaining form factors are, will be presented in our future work \cite{Hermsen:2025}.
With such a representation the evaluation of loop corrections is reduced to the consideration of scalar functions, avoiding the somewhat more tedious tensor loop integrals. \\\par
\begin{figure*}[!b]
\centering
    \subfloat{
        \centering
        \includegraphics[width=.22\textwidth]{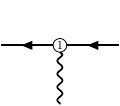}
    }
    ~ 
    \subfloat{
        \centering
        \includegraphics[width=.22\textwidth]{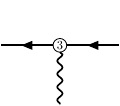}
    }
     ~ 
    \subfloat{
        \centering
        \includegraphics[width=.22\textwidth]{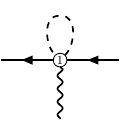}
    }
    \\[1em]
    \subfloat{
        \centering
        \includegraphics[width=.22\textwidth]{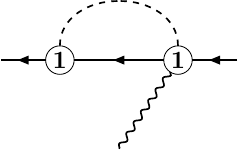}
    }
    ~
   \subfloat{
        \centering
        \includegraphics[width=.22\textwidth]{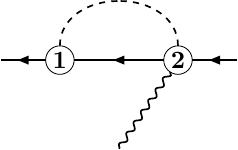}
    }
    ~\subfloat{
        \centering
        \includegraphics[width=.22\textwidth]{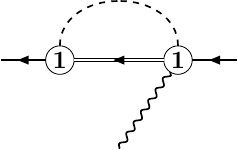}
    }
    ~
   \subfloat{
        \centering
        \includegraphics[width=.22\textwidth]{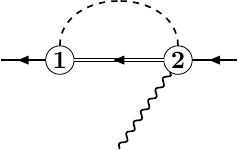}
    }
    \\[1em]
   \subfloat{
        \centering
        \includegraphics[width=.22\textwidth]{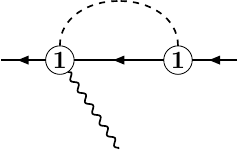}
    }
    ~
    \subfloat{
        \centering
        \includegraphics[width=.22\textwidth]{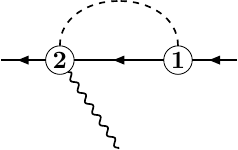}
    }
    ~\subfloat{
        \centering
        \includegraphics[width=.22\textwidth]{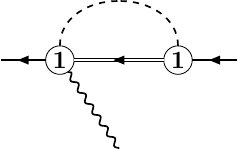}
    }
    ~
    \subfloat{
        \centering
        \includegraphics[width=.22\textwidth]{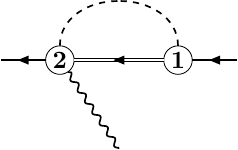}
    }
    \\[1em]
    \subfloat{
        \centering
        \includegraphics[width=.22\textwidth]{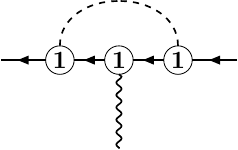}
    }
    ~
    \subfloat{
        \centering
        \includegraphics[width=.22\textwidth]{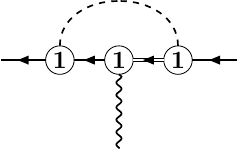}
    }
    ~
    \subfloat{
        \centering
        \includegraphics[width=.22\textwidth]{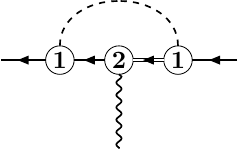}
    }
    \\[1em]
   \subfloat{
        \centering
        \includegraphics[width=.22\textwidth]{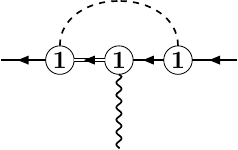}
    }
    ~
    \subfloat{
        \centering
        \includegraphics[width=.22\textwidth]{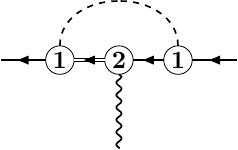}
   }
    ~
    \subfloat{
        \centering
        \includegraphics[width=.22\textwidth]{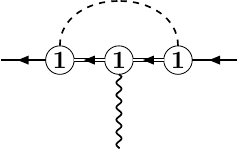}
    }
    \caption{The tree and one-loop diagrams contributing to the axial-vector form factor of the nucleon up to chiral order $Q^4$ as given in Eq. \eqref{res-FA}. The  wavy, dotted solid, and double lines represent the axial current, pion, nucleon, and $\Delta$-isobar, respectively. The chiral order of all vertices is indicated.}
    \label{fig:treeloopdiagrams}
\end{figure*}

In this work we apply the chiral Lagrangian with nucleon and $\Delta$-isobar degrees of freedom as developed in Refs. \cite{Bernard:1993bq,Bernard:1998gv,Fearing:1997dp,Schindler:2006it, Fuchs:2003qc, Chen:2012nx, Ando:2006xy,Hemmert:2003cb,Procura:2006gq,Yao:2017fym, Ellis:1997kc,Lutz:2020dfi}. Although early work used heavy-baryon ChPT \cite{Bernard:1993bq,Bernard:1998gv,Fearing:1997dp}, later the relativistic form of the chiral Lagrangian was advocated \cite{Schindler:2006it,Fuchs:2003qc,Chen:2012nx,Ando:2006xy}. The $\Delta$-isobar degrees of freedom were considered so far only in Refs. \cite{Hemmert:2003cb,Procura:2006gq,Yao:2017fym,Ellis:1997kc,Lutz:2020dfi}. We will use the conventions of Refs. \cite{Lutz:2020dfi,Sauerwein:2021jxb}, in which a renormalization scheme based on the Passarino-Veltman reduction scheme \cite{Passarino:1978jh} was applied. Our starting point is then
\begin{eqnarray}
&& F_A^\mu(\bar p,p) = \Big(  \frac{1}{\sqrt{2}} \,g_A\, \gamma^\nu \,\gamma_5 \Big)
\Big(Z_N\, g_{\nu}^{\;\mu} - \frac{q_\nu \,q^\mu}{q^2 -m_\pi^2}\,\Big[ Z_N+f_\pi/f +Z_\pi - 2\Big] \Big)
\nonumber\\
&& \hspace{2em}
+ \, \frac{1}{\sqrt{2}}\,\Big(
g_R \, q^2 +4 \, g^+_\chi \,( 2\,B_0\,m ) \Big) \,\gamma^\mu \, \gamma_5 - \frac{2\,M}{\sqrt{2}}\,\Big( 
4\,g_\chi ^++ g_\chi^- \Big)\,\gamma_5\,\frac{q^\mu \,(2\,B_0\,m)}{q^2-m_\pi^2} 
\notag \\ && \hspace{2em}
- \, \frac{1}{\sqrt{2}}\,g_R\,\slashed{q}\,\gamma_5\,q^\mu
+\frac{g_A}{f^2}\,  \Big\{J^\mu_{\pi}(\bar p, p) + J_{ N \pi }^\mu(\bar p, p)   +  J_{\pi N}^\mu(\bar p, p) \Big\}
+ \frac{g_A^3}{4\,f^2}\,J^\mu_{N\pi N}(\bar p, p)  
\notag \\ && \hspace{2em}
+\, \frac{f_S}{3\,f^2}\,  \Big\{ J_{ \Delta \pi }^\mu(\bar p, p) + J_{\pi \Delta}^\mu (\bar p, p)  \Big\} + 
 \frac{2\,g_A\,f_S}{3\,f^2}\,\Big\{J^\mu_{N\pi \Delta }(\bar p, p) + J^\mu_{\Delta \pi N }(\bar p, p) \Big\} 
\notag \\ && \hspace{2em}
+\, \frac{5\,h_A\,f_S^2}{9\,f^2}\,J^\mu_{\Delta \pi \Delta}(\bar p, p) 
+{\mathcal O} \big( Q^4\big)\,,
 \label{res-FA}
 \end{eqnarray}
in terms of the relevant low-energy constants (LECs), the nucleon wave-function factor $Z_N$, the pion wave-function factor $Z_\pi$, the pion decay constant $f_\pi$ and the loop integrals $J^\mu_{\pi},\,J^\mu_{L\pi},\,J^\mu_{\pi R}$, and $J^\mu_{L\pi R}$ with $L,R\in\{N,\Delta\}$. Since we focus on the axial-vector form factor $G_A$ we do not display $f_\pi$ and $Z_\pi$, since they only contribute to the pseudoscalar form factor $G_P$ \cite{Hermsen:2024eth,Schindler:2006it}. The LECs $f,M$ are the pion decay constant and the mass of the nucleon in the chiral limit respectively. In addition to the leading-order LECs $g_A, h_A, f_S$ we also need the subleading-order LECs $g_R, g_\chi^{\pm}$ in Eq. (\ref{res-FA}).  The LECs in our current study are taken from Ref. \cite{Hermsen:2024eth} and recalled in Table \ref{tab:listLECs}. For the detailed display of the respective Lagrangian we refer the reader to the Refs. \cite{Lutz:2020dfi,Sauerwein:2021jxb,Heo:2022huf}.
The tree and one-loop diagrams implied by Eq. \eqref{res-FA}, that contribute to the axial-vector form factor are shown in Fig. \ref{fig:treeloopdiagrams}.\\

\par

\begin{table*}[t]
\centering
\begin{tabular}{lrp{10pt}lr}
\toprule
LEC & value & & LEC & value \\ 
\midrule
  $f$ [MeV] & 83.43  && $f_S$ & 1.5857\\
$M$ [MeV] & 893.79 && $g_S$ [GeV$^{-1}$] & 0.9163\\
$\Delta$ [MeV] & 306.63 && $g_T$ [GeV$^{-1}$] & 1.5035\\ 
$g_A$ & 1.1449  && $f_E$ [GeV$^{-1}$] & 0.6949
\\\bottomrule\\
\end{tabular}
\caption{Values of LECs from Ref. \cite{Hermsen:2024eth}.}
\label{tab:listLECs}
\end{table*}

In $d$ dimensions, the loop integrals are given by
\begin{equation}
J^\mu_{\cdots} (\bar p, p) =  i\,\mu^{4-d}
\sum_{\vec{n}\in\mathbb{Z}^3}\,
\int \frac{d^d l}{(2\pi)^{d}} \,
e^{i\, \vec{l}\,\cdot \, \vec{x}_n}\,
K^\nu_{\cdots} (\bar p, p; l)\,\Big(  \delta_{\nu}^{\;\mu} - q_\nu \,q^\mu/(q^2-m_\pi^2) \Big) \ ,
\label{def-Js}
\end{equation}
where we recall $\vec x_n = L\,\vec n$, $l$ is the loop momentum and $\cdots\in\{\pi,L\pi,\pi R, L \pi R\}$.

The sum is split into two parts: Its infinite-box limit, $L \to \infty$, corresponds to the $\vec n =(0,0,0)$ term. All other terms in the sum are finite-volume corrections. While the infinite-volume term requires dimensional regularization together with a suitable renormalization procedure, the volume correction term is finite and can be computed at $d=4$. \\\par

From our previous works \cite{Lutz:2020dfi,Sauerwein:2021jxb,Isken:2023xfo} we recall the expressions for the functions $K^\mu_{\cdots} (\bar p, p; l)$:
\begin{eqnarray}
&&K_\pi^\mu = - \frac{1}{\sqrt{2}} \,
\frac{ \gamma^{\mu} \, \gamma_5}{l^2-m_\pi^2} 
 + \frac{1}{\sqrt{2}} \,\frac{q^\mu }{3\,m_\pi^2} \,\frac{4\,M_N}{l^2-m_\pi^2}\, \gamma_5\,, 
\nonumber\\
&&K^\mu_{ N \pi} =  -\frac{1}{\sqrt{2}}\,\frac{\slashed{l}\,\gamma_5}{l^2-m_\pi^2} \, S_N(\bar{p}-l) \, \Big( \gamma^{\mu}  + 2\,g_S\,l^\mu 
- 2\,g_T\, i \, \sigma^{\mu\nu} \, l_{\nu} 
\nonumber \\ &&\hspace{3.4em} 
+\,\frac{1}{2}\,g_V \,
\big[ \gamma^{\mu} \, p\cdot l+\slashed{l}\, p^{\mu} + \gamma^{\mu} \, (\bar{p}-l)\cdot l+\slashed{l}\,(\bar{p}-l)^{\mu}\big]+ 4\,g_F\, i \, \sigma^{\mu\nu} \, q_{\nu} \Big) \,, 
\nonumber \\
&&K^\mu_{ \pi N} =  -\frac{1}{\sqrt{2}}\,\Big( \gamma^{\mu}  + 2\,g_S\,l^\mu 
+ 2\,g_T\, i \, \sigma^{\mu\nu} \, l_{\nu} + 4\,g_F\, i \, \sigma^{\mu\nu} \, q_{\nu} 
\nonumber \\ &&\hspace{3.4em} 
+\,\frac{1}{2}\,g_V \,
\big[ \gamma^{\mu} \, {\bar p}\cdot l+\slashed{l}\, {\bar p}^{\mu} + \gamma^{\mu} \, (p-l)\cdot l+\slashed{l}\,(p-l)^{\mu}\big] \Big)\, S_N(p-l) \,\frac{\slashed{l}\,\gamma_5}{l^2-m_\pi^2} \,, 
\nonumber \\
&&K^\mu_{N \pi N} = - \frac{1}{\sqrt{2}} \, \slashed{l}  \gamma_5  \, S_N(\bar{p}-l) \,\frac{ \gamma^{\mu}  \gamma_5}{l^2-m_\pi^2} \, S_N(p-l) \, \slashed{l} \gamma_5 \, , 
\nonumber \\
&&K^\mu_{\Delta \pi} =  \frac{1}{\sqrt{2}}\,
\Big( (f_A^- 
-5 \,f_A^+ )\, S_{\Delta}^{\nu\mu}(\bar{p}-l) \, \slashed{l}- (f_A^ -+5\,f_A^+)\,S_\Delta^{\nu\alpha}(\bar{p}-l) \, l_{\alpha} \, \gamma^\mu
\nonumber \\ &&\hspace{3.4em}
- \,4 \, f_M \, \big[S_{\Delta}^{\nu\alpha}(\bar{p}-l) \, \gamma^\mu-S_{\Delta}^{\nu\mu}(\bar{p}-l) \, \gamma^\alpha\big] \, q_\alpha
\Big) \,
\frac{l_\nu \, \gamma_5}{l^2-m_\pi^2} \,, 
\nonumber \\
&&K^\mu_{\pi \Delta} =  -\frac{1}{\sqrt{2}}\,
 \frac{\gamma_5 \,l_\nu}{l^2-m_\pi^2}
 \Big( (f_A^- -5\,f_A^+ )\, \slashed{l} \, S_{\Delta}^{\mu\nu}(p-l)
 -(f_A^-+5\,f_A^+)\,\gamma^\mu \, l_\alpha \, S_{\Delta}^{\alpha\nu}(p-l)
\nonumber \\ && \hspace{3.4em}
-\, 4 \, f_M \, q_\alpha \, \big[\gamma^{\alpha} \, S_{\Delta}^{\mu\nu}(p-l)-\gamma^\mu \, S_{\Delta}^{\alpha\nu}(p-l)\big]\Big) \,, 
\nonumber\\
&&K^\mu_{N\pi \Delta} =  \frac{1}{\sqrt{2}}\,\frac{\slashed{l}\gamma_5 \,l_\tau}{l^2-m_\pi^2}\,S_N(\bar{p}-l)
\nonumber \\ && \hspace{6.4em}\times \,
\Big(f_S\, S_{\Delta}^{\mu\tau}(p-l) + 2\,f_E\big[\gamma_{\mu}S_{\Delta}^{\alpha\tau}(p-l)-\gamma^\alpha S_{\Delta}^{\mu\tau}(p-l)\big]\,q_\alpha\Big) \,, 
\nonumber \\
&&K^\mu_{\Delta \pi N} =  \frac{1}{\sqrt{2}}\,
\Big(f_S\, S_{\Delta}^{\sigma\mu}(\bar{p}-l) + 2\,f_E\, \big[S_{\Delta}^{\sigma\mu}(\bar{p}-l) \gamma^{\alpha}-S_{\Delta}^{\sigma\alpha}(\bar{p}-l) \gamma^\mu\big] q_\alpha\Big) 
\nonumber \\ && \hspace{6.4em}\times \,
 S_N(p-l)\frac{l_\sigma\,\slashed{l}\gamma_5}{l^2-m_\pi^2} \,, 
\nonumber \\
&&K^\mu_{\Delta \pi \Delta} =  - \frac{1}{\sqrt{2}}\,l_{\sigma}\,S_\Delta^{\sigma\alpha}(\bar{p}-l) \,\frac{\gamma^\mu\,\gamma_5}{l^2-m_\pi^2}\, S^\Delta_{\alpha\beta}(p-l) \,l^{\beta}\,,
\label{def-Ks}
\end{eqnarray}
with the baryon propagators \cite{Lutz:2001yb,Semke:2005sn,Haberzettl:1998rw}
\begin{eqnarray}
&&S_{N}(k) = \frac{1}{\slashed{k}-M_N }\,,
 \nonumber\\
&&S_{\Delta}^{\mu\nu}(k) =\frac{-\,1}{\slashed{k}-M_{\Delta}}
\bigg(g^{\mu\nu}-\frac{\gamma^{\mu}
\gamma^{\nu}}{d-1}+\frac{k^{\mu}\gamma^{\nu}-k^{\nu}\gamma^{\mu}}{(d-1)\,M_{\Delta}}-
\frac{(d-2)\,k^{\mu}k^{\nu}}{(d-1)\,M_{\Delta}^2}\bigg)\,,
\label{def-propagtors}
\end{eqnarray}
the two-body LECs $g_S,g_V, g_T, f_\pm,\,f_M, g_F$, and the subleading axial $\Delta N$ transition LEC $f_E$.
Our main goal here is the derivation of the finite-box effects in the loop integrals containing the functions of Eq. (\ref{def-Ks}).  
In our previous works we considered in detail the infinite-volume limit thereof, including the expansion in chiral powers according to a power-counting scheme formulated in terms of on-shell hadron masses \cite{Lutz:2020dfi,Sauerwein:2021jxb}. \\\par

A few comments may be useful here. While replacing free-space hadron masses by their in-box correspondents for stable states is straight forward, matters turn more intricate for unstable states like the isobar baryon in Eq. (\ref{def-propagtors}). In the infinite volume the physical properties of such a resonance may be taken justice to by a spectral weight with
\begin{eqnarray}
    S_{\Delta}^{\mu\nu}(k) \to \int^\infty_{m_\pi + M_N} d M \, \rho_{\Delta }(M )\,S_{\Delta}^{\mu\nu}(k, M) \,.
\end{eqnarray}
The frequently assumed quasi-particle approximation is given with
$\rho_\Delta(M) = \delta (M- M_\Delta)$
with an effective mass parameter $M_\Delta$ of the resonance state. More refined approximations can be generated by using a finite sum of discrete states of masses $M_\Delta^{(n)}$ constructed to provide an envelop of the physical spectral mass distribution of a resonance. Here one would insist on  $M_\Delta^{(n)} < M_\Delta +  \Gamma_\Delta $ with $\Gamma_\Delta $ the width  of that resonance. Such a strategy is closely linked to what happens in a finite box. Depending on the box size there are a finite number of discrete energy levels only that one may  associate with that resonance state. Moreover, unlike for a stable state the dependence of the mass parameters $M_\Delta^{(n)}$ do not show only an exponentially suppressed volume dependence, but also L\" uscher like terms that are polynomial in $1/L$ \cite{Luscher:1986pf}.

\section{A new set of in-box basis loop functions}
\label{sec:Basis}
In this Section we will express the loop integrals introduced in Eq. (\ref{def-Ks}) in terms of a new set of basis functions that generalize the Passarino-Veltman reduction scheme \cite{Passarino:1978jh,Isken:2023xfo} to the in-box case. The first step is to apply the projected form of Eq. (\ref{ProjAxialcurrent}) that avoids the need to consider tensor-type loop integrals. We write
\begin{eqnarray}
&& G_A(\bar{p},p)=
 g_A\,Z_N+ 4\,g^+_\chi \,m_\pi^2  + g_R \,t
+\,\frac{g_A}{f^2}\,  \Big\{J^A_{\pi}(\bar{p},p) + J_{ N \pi }^A(\bar{p},p) +  J_{\pi N}^A(\bar{p},p)\Big\}
\nonumber\\ && \hspace{2em}
+\, \frac{g_A^3}{4\,f^2}\,J^A_{N\pi N}(\bar{p},p) 
+ \frac{f_S}{3\,f^2}\,  \Big\{J_{ \Delta \pi }^A(\bar{p},p) +  J_{\pi \Delta}^A (\bar{p},p)\Big\}  
 \nonumber\\ && \hspace{2em}
+ \, \frac{2\,g_A\,f_S}{3\,f^2}\,\Big\{J^A_{N\pi \Delta }(\bar{p},p)+ J^A_{\Delta \pi N }(\bar{p},p)\Big\}
+ \frac{5\,h_A\,f_S^2}{9\,f^2}\,J^A_{\Delta \pi \Delta}(\bar{p},p)
 + {\mathcal O} \big( Q^4\big) \,,
\label{res-GA}
\end{eqnarray}
where for any loop function we introduced its 
projected form, indicated by the replacement $J^\mu_{\cdots}(\bar p, p) \to J^A_{\cdots}(\bar p, p)$. Since the in-box form factors depend on both momenta $\vec {\bar p}$ and $\vec p$, we keep the notation $J^A_{\cdots}(\bar p, p)$.
\\\par
To present the form factor in a compact form, we will construct a convenient set of renormalized basis functions, with corresponding kinematical coefficient functions \cite{Hermsen:2024eth,Lutz:2020dfi}.
Our truncated expressions imply an expansion of the coefficient functions in chiral orders, for which we use the power counting
\begin{eqnarray}
t\sim  m_\pi^2 \sim Q^2 \,,\qquad \delta= M_\Delta - M_N \,\Big( 1+ \frac{\Delta}{M} \Big) \sim Q^2 \,,
\label{def-counting-rule}
\end{eqnarray}
where we keep the on-shell masses unexpanded.  The masses of the nucleon and the $\Delta$-isobar in the chiral limit are denoted by $M$ and $M+ \Delta$, respectively. 
The basis functions are assigned specific chiral orders as suggested by their leading order in a strict chiral expansion. We expect that the chiral Ward identities hold for such truncated expressions \cite{Lutz:2020dfi,Sauerwein:2021jxb}.
\\\par
It is sufficient to consider only scalar loop integrals. We introduce an intermediate, over-complete set of basis functions by the integrals 
\begin{eqnarray}
J^{(\bar{a} \IS h \IS a)}_{\pi} (\bar p,p) &=& i\,\mu^{4-d}\,
\sum_{\vec{n}\in\mathbb{Z}^3} \int \frac{d^d l}{(2\pi)^{d}} \, \frac{e^{i\, \vec{l}\,\cdot\, \vec{x}_n}\,(l^2)^h}{l^2-m_\pi^2}
(\bar p \cdot l)^{\bar{a}}  \,(l \cdot p)^{a}\,, 
\nonumber \\
J^{(\bar{a} \IS h \IS a)}_{L \pi}(\bar p,p) &=& -i\,\mu^{4-d}\,
\sum_{\vec{n}\in\mathbb{Z}^3} \int \frac{d^d l}{(2\pi)^{d}} \, \frac{e^{i\, \vec{l}\,\cdot\, \vec{x}_n}\,(l^2)^h}{l^2-m_\pi^2}
\frac{(\bar p \cdot l)^{\bar{a}}  \,(l \cdot p)^{a}}{(l-\bar p)^2- M_L^2}\,, 
\nonumber \\
J^{(\bar{a} \IS h \IS a)}_{\pi R} (\bar p,p) &=& -i\,\mu^{4-d}\,
\sum_{\vec{n}\in\mathbb{Z}^3} \int \frac{d^d l}{(2\pi)^{d}} \, \frac{e^{i\, \vec{l}\,\cdot\, \vec{x}_n}\,(l^2)^h}{l^2-m_\pi^2}
\frac{(\bar p \cdot l)^{\bar{a}}  \,(l \cdot p)^{a}}{(l-p)^2- M_R^2}\,,
\nonumber \\
J^{(\bar{a} \IS h \IS a)}_{L \pi R}(\bar p,p) &=& i\,\mu^{4-d}\,
\sum_{\vec{n}\in\mathbb{Z}^3} \int \frac{d^d l}{(2\pi)^{d}} \, \frac{e^{i\, \vec{l}\,\cdot\, \vec{x}_n}\,(l^2)^h}{l^2-m_\pi^2}
\frac{(\bar p \cdot l)^{\bar{a}} }{(l-\bar p)^2- M_L^2}\,\frac{(l \cdot p)^{a}}{(l-p)^2-M_R^2}\,,
\label{def-overcomplete-basis}
\end{eqnarray}
where $L,R$  either refer to a nucleon or a $\Delta$-isobar field.
An additional tadpole structure, $J_{L/R}^{(\bar{a} \IS h \IS a)}(\bar p, p)$ with a heavy field with $M_{L/R}$ only, can be included, but in our renormalization scheme such terms will be dropped. 
\\
\par
The contributions of any of the loop functions in Eq. (\ref{res-FA}) to the axial-vector form factor $G_A( \bar p, p)$ are derived by performing the trace in Eq. (\ref{ProjAxialcurrent}). For instance, this leads to
\begin{eqnarray}
&& J_{N\pi}^{A}(\bar p, p) = 
2\,(1+ 8\,g_F\,M_N)\,J_{N\pi}^{(1 \IS 0 \IS 0)} 
+\frac{4\,(1+8\,g_F\,M_N)\,M_N^2}{t-4\,M_N^2}\,\big(J_{N\pi}^{(1 \IS 0 \IS 0)} + J_{N\pi}^{(0 \IS0 \IS1)}\big)
\nonumber\\
&& \hspace{.5em}
- \,J_{N\pi}^{(0 \IS 1 \IS 0)}
+ g_V\,\big(J_{N\pi}^{(2 \IS 0 \IS 0)} + J_{N\pi}^{(1 \IS 0 \IS 1)}\big)
+ \frac{4\,\big(g_S-(d-3)\,g_T\big)\,M_N}{d-2}\,J_{N\pi}^{(0 \IS 1 \IS 0)}
\nonumber \\  && \hspace{.5em}
-\, \frac{(3\,d-4)\,g_V}{2\,(d-2)}\,J_{N\pi}^{(1 \IS 1 \IS 0)}
- \frac{g_V}{2}\,J_{N\pi}^{(0 \IS 1 \IS 1)}
+\frac{(d-1)\,g_V}{2\,(d-2)}\,J_{N\pi}^{(0 \IS 2 \IS 0)}
-\frac{4\,(g_S+g_T)\,M_N}{(d-2)\,t}
\,\Big(J_{N\pi}^{(2 \IS 0 \IS 0)} 
\nonumber \\  &&\hspace{1em}
- \, 2\,J_{N\pi}^{(1 \IS 0 \IS 1)}+J_{N\pi}^{(0 \IS 0 \IS 2)}\Big)
+ \frac{g_V}{(d-2)\,t}\,\Big(J_{N\pi}^{(3 \IS 0 \IS 0)}-2\,J_{N\pi}^{(2 \IS 0 \IS 1)} + J_{N\pi}^{(1 \IS 0 \IS 2)}\Big)
\nonumber\\ &&\hspace{.5em}
- \, \frac{g_V}{2\,(d-2)\,t}\,\Big(J_{N\pi}^{(2 \IS 1 \IS 0)} - 2\,J_{N\pi}^{(1 \IS 1 \IS 1)} + J_{N\pi}^{(0 \IS 1 \IS 2)}\Big)
+ \frac{1}{t-4\,M_N^2}\,\bigg\{ \Big[2\,g_V\,M_N^2
\nonumber\\ &&\hspace{1em}
+\,\frac{4\,\big(g_S-(2\,d-5)\,g_T\big)\,M_N}{d-2} \Big] J_{N\pi}^{(2 \IS 0 \IS 0)}
+ \Big[4\,g_V\,M_N^2+\frac{8\,\big(g_S-(d-3)\,g_T\big)\,M_N}{d-2}\Big] J_{N\pi}^{(1 \IS 0 \IS 1)}
\nonumber\\ &&\hspace{.5em}
+\,\Big[2\,g_V\,M_N^2+\frac{4\,(g_S+g_T)\,M_N}{d-2}\Big] J_{N\pi}^{(0 \IS 0 \IS 2)}
+ 2\,M_N\,(2\,g_T-g_V\,M_N)\,\Big(J_{N\pi}^{(1\IS 1 \IS 0)} + J_{N\pi}^{(0 \IS 1 \IS 1 )}\Big)
\nonumber\\ &&\hspace{.5em}
- \, \frac{g_V}{d-2}\,\Big(J_{N\pi}^{(3 \IS 0 \IS 0)} + 2\,J_{N\pi}^{(2\IS 0\IS 1)} + J_{N\pi}^{(1 \IS 0 \IS 2)}\Big)
+\frac{g_V}{2\,(d-2)}\,\Big(J_{N\pi}^{(2 \IS 1 \IS 0)} + 2\,J_{N\pi}^{(1 \IS 1 \IS 1)}+J_{N\pi}^{(0 \IS 1 \IS 2)}\Big) \bigg\} \,, 
\nonumber\\
&&J^A_{N\pi N}(\bar p, p) = -
\,4\,J_{N\pi N}^{(1 \IS 0 \IS 1)} 
+2\,J_{N\pi N}^{(1 \IS 1 \IS 0)}
+2\,J_{N\pi N}^{(0 \IS 1 \IS 1)}
-J_{N\pi N}^{(0 \IS 2 \IS 0)} - \frac{4\,(d-4)\,M_N^2}{d-2}\,J_{N\pi N}^{(0 \IS 1 \IS 0)}
\nonumber\\ &&\hspace{.5em}
- \, \frac{8\,(d-3)\,M_N^2}{(d-2)\,(t-4\,M_N^2)}\,\Big(J_{N\pi N}^{(2 \IS 0 \IS 0)} + 2\,J_{N\pi N}^{(1 \IS 0 \IS 1)} + J_{N\pi N}^{(0 \IS 0 \IS 2)}\Big)
+ \frac{8\,M_N^2}{t-4\,M_N^2}\,\Big(J_{N\pi N}^{(1 \IS 1 \IS 0)} + J_{N\pi N}^{(0 \IS 1 \IS 1)}\Big)
\nonumber\\ &&\hspace{.5em}
- \,\frac{8\,M_N^2}{(d-2)\,t}\,\Big(J_{N\pi N}^{(2 \IS 0 \IS 0)} - 2\,J_{N\pi N}^{(1 \IS 0 \IS 1)} + J_{N\pi N}^{(0 \IS 0 \IS 2)}\Big) \,.
\end{eqnarray}
While the over-complete set of basis functions result is a useful 
first reduction of the form factor, those basis functions are not independent, and therefore lead to different obstacles: Due to our projection technique, artificial singularities occur at $t=0$, which only get canceled in an appropriate minimal set of basis integrals, as we already showed in the infinite-volume limit \cite{Hermsen:2024eth,Lutz:2020dfi,Sauerwein:2021jxb}. Additional each element of the over-complete set would need to be treated in dimensional regularization separately, due to there momentum dependency. Consequently, each element  can contain power-counting violating terms. Their identification would be laborious and the determination of the needed subtractions not as straightforward as in the basis functions. Lastly, these results would not be efficient in numerical applications, due to the large amount of elements and their dependencies on each other. 
\\
\par
Restricting ourself to the infinite-volume limit, there are, according to the Passarino--Veltman scheme, only three independent scalar loop functions needed, of tadpole-, bubble-, and triangle-type. (An additional structure is needed in cases of artificial singularities, as shown in Ref. \cite{Lutz:2020dfi}.)
The number of elements in the set of basis functions grows in the finite box, as will be clear from the rest of this Section. While we still have tadpole-, bubble-, and triangle-loop functions, they now depend on three different indices, due to the broken rotational and translational symmetries in the finite box and the explicit dependence on the incoming and outgoing three momenta.
\\
\par
In Appendix \ref{app:redInBox}  we first provide our minimal set of basis functions for the loop-functions introduced in Eq. (9). In the main text  we only provide their renormalized forms, which in our notation come with an extra bar. We start with our tadpole basis functions,
\begin{eqnarray}
&&\bar I^{(\bar a , \, h,  \,a)}_{\pi}(\vec{ \bar p}, \,\vec{ p}\,) = \delta_{\bar a,0}\,\delta_{h,0}\,\delta_{a,0}\, \bar I_\pi
+ \,
\sum^{\vec n \neq 0}_{\vec{n}\in\mathbb{Z}^3}\,X_\pi^{(\bar{a} + 2\,h + a)}[\vec{x}_n]\,\big(\vec{\bar p}\cdot \vec{x}_n \big)^{\bar a}\,\big(\vec{x}_n^{\,2}\big)^h\, \big( \vec{p}\cdot \vec{x}_n \big)^a
\ ,\nonumber \\
&&\bar  I_\pi = \frac{m_\pi^{2}}{16 \pi^2}\,\log\frac{m_\pi^2}{\mu^2} \sim Q^{2}
\,, \nonumber \\ 
&&X_\pi^{(n_x)}[\,\vec{x}_n] = \frac{(-1)^{n_x /2}}{4\,\pi^2}\,K_{1 + n_x } \big[\,m^2_\pi,\, \vert \vec{x}_n\vert\, \big] \,, \nonumber \\
&&K_{n} [\,m^2, \,x ] = \bigg( \frac{m }{x} \bigg)^n
K_n (\,m \, x\,) \,,
\label{def-tadpole-basis}
\end{eqnarray}
where $p^2_0= M_N^2+\vec{ p\,}^2$,
$\bar p^2_0= M_N^2+\vec{\bar  p}\,^2$, and $K_n(m\,x)$ denotes the modified Bessel function. Note that $\bar I^{(\bar a,\,h , \,a)}_{\pi}(\vec{ \bar p},\, \vec{p}\,) = 0$ if $\bar{a}+a$ odd. The chiral power of such renormalized tadpole terms is readily identified with $m_\pi\sim Q$. The infinite-volume part $\bar I_\pi$ scales with $Q^{2}$. If we assign the typical value $x_n \sim 1/m_\pi$ together with $\vec p^{\,2 } \sim \vec {\bar p}^{\,\,2} \sim m^2_\pi$ we find a universal scaling of both parts, with  $\bar{I}_\pi^{(\bar{a},\,h,\,a)}\sim Q^{2+2\,\bar{a} + 2\,h + 2\,a}$. In turn we may decompose the renormalized form of $J^A_\pi(\bar{p},p)$, for which we find
\begin{eqnarray}
\bar  J^A_\pi(\bar p, p) = - \,\bar I^{(0 , 0,  0)}_{\pi}(\vec{ \bar p},\, \vec{ p}\,) \,.
\label{eqn:resJAPi}
\end{eqnarray}
The infinite-volume part of Eq. \eqref{eqn:resJAPi}  confirms our previous findings \cite{Hermsen:2024eth,Lutz:2020dfi}. 
In Ref. \cite{Lutz:2014oxa}, where we studied the finite-volume effects of the baryon masses, the tadpole basis was defined as 
\begin{equation}
I_\pi^{(a)} = i\, \mu^{4-d}\,\sum_{\vec{n}\in\mathbb{Z}^3}\,\int\frac{\dd[d]{l}}{(2\,\pi)^d}\frac{e^{i\,\vec{l}\,\cdot\,\vec{x}_n}\,l_0^a}{l^2-m_\pi^2}\,,
\label{eqn:matthias-basis-definition}
\end{equation}
because only the kinematic point with $\vec p = \vec{\bar p}=0$ was considered. Since we insist that Eq. \eqref{def-tadpole-basis} forms a minimal set of basis functions, it is possible to express each element from Eq. \eqref{eqn:matthias-basis-definition} as a linear combination thereof. While for $a$ odd the integrals in \eqref{eqn:matthias-basis-definition} vanish, for even $a$ we obtain in dimensional regularization
 \begin{eqnarray}
&&I_\pi^{(2\,a)} = (2\,a-1)!!\,\Bigg\{\Bigg[\prod_{j=0}^{a-1}\frac{1}{d + 2\,j}\Bigg]\,i\,\int\frac{\dd[d]{l}}{(2\,\pi)^d}\frac{\mu^{4-d}\,\big(l^2\big)^{a}}{l^2-m_\pi^2}
\nonumber \\ &&\hspace{10em} 
 + \, (-1)^{a}\,\frac{1}{2^{2-a}\,\pi^{2}}\,\sum_{\vec{n}\in\mathbb{Z}^3}^{\vec{n}\neq \vec{0}}
K_{1+a}\big[\,m^2_\pi,\, \vert \vec{x}_n\vert\, \big] \Bigg\}\,.
\label{eqn:matthias-basis-calculated}
\end{eqnarray}
With Eq. \eqref{eqn:matthias-basis-calculated} we recover Eqs. (2) and (19) from Ref. \cite{Lutz:2014oxa}, where only the cases $a\in\{0,2\}$ were needed.  We identify the combination in the curly brackets as $I_{\pi,a}^{(0,\,0,\,0)}$ defined in Eq. \eqref{def-intermidate-feynman}. By applying Eq. \eqref{app:eqn:iterationTadpole} iteratively, we find 
\begin{eqnarray}
&&I_\pi^{(2\,a)} = m_\pi^{2\,a}\Bigg[\prod_{j=0}^{a-1}\,\frac{1}{d+2\,j}\Bigg]\,I_\pi^{(0,0,0)}
+ \sum_{n=1}^a\,\Bigg[\prod_{j=0}^{n-1}\,\frac{1}{d+2\,(n-1+j)}\Bigg]
\nonumber \\ &&\hspace{8em}\times\, 
\Bigg[\prod_{j=0}^{a-n-1}\,\frac{1}{d+4\,n+2\,j}\Bigg]\,(-1)^n\,\binom{a}{n}\,m_\pi^{2\,(a-n)}\,I_\pi^{(0,n,0)}\,.
\end{eqnarray}
With this we indeed proved the equivalence of both sets of basis elements in the case of $\vec{p}=\vec{ \bar p} = \vec{0}$. We give two explicit examples in terms of renormalized elements:
\begin{eqnarray}
\bar{I}_\pi = \bar{I}_\pi^{(0,0,0)}\,,\qquad
\bar{I}_\pi^{(2)} = \frac{m_\pi^2}{4}\,\bar{I}_\pi^{(0,0,0)} - \frac{1}{4}\,\bar{I}_\pi^{(0,1,0)}\,.
\end{eqnarray}

 We continue with the bubble-loop integrals, for which we identified the following set of basis functions:
\begin{eqnarray}
&&\bar I^{(\bar a,\, h , \,a)}_{L\pi}(\vec{ \bar p}, \vec{ p}\,) = \delta_{\bar a,0}\,\delta_{ h,0}\,\delta_{ a,0}\, \bar I_{ L \pi}
+ 
\sum^{\vec n \neq 0}_{\vec{n}\in\mathbb{Z}^3} \,X_{L\pi}^{(\bar{a} + 2\,h +  a)}[\,\vec{\bar{p}}, \, \vec{x}_n\,]\,
 \big(\vec{\bar p}\cdot \vec{x}_n \big)^{\bar{a}}\, \big(\vec{x}_n^{\,2}\big)^h\, \big(\vec{p}\cdot\vec{x}_n \big)^a
 \,, \nonumber \\
&&\bar{I}_{L\pi} = \frac{\gamma_N^L-2}{16\,\pi^2} - \frac{1}{16\,\pi^2}\,\int_0^1\,\dd{z} \,\log\,\frac{F_{L\pi}(z)}{M_L^2} \sim Q
\,, \nonumber \\
&&X_{L\pi}^{(n_x)}[\,\vec{\bar{p}},\, \vec{x}_n\,] =  \frac{(-1)^{\lfloor (1-n_x ) / 2 \rfloor}}{8\,\pi^2}\,\int_0^1 d\,z\,K_{n_x}\big[\, F_{L\pi}(z),\, \vert \vec{x}_n \vert \big]\,
 {\rm cs}_{n_x}\big( z\,\vec{ \bar p}\cdot \vec{x}_n \big)
 \,, \nonumber \\
&&F_{L\pi}(z) = z\,M_L^2 + (1-z)\,m_\pi^2- (1-z)\,z\,\bar p^2
\,,\nonumber \\
&&\bar I^{(\bar a,\, h , \,a)}_{\pi R}(\vec{ \bar p}, \vec{ p}\,) = \delta_{\bar a,0}\,\delta_{ h,0}\,\delta_{ a,0}\, \bar I_{\pi R}
+ \sum^{\vec n \neq 0}_{\vec{n}\in\mathbb{Z}^3} \,X_{\pi R}^{(\bar{a} + 2\,h + a)}[\,\vec{x}_n,\,\vec{p}\,]\,
 \big(\vec{\bar p}\cdot \vec{x}_n \big)^{\bar a}\,\big(\vec{x}_n^{\,2}\big)^h\, \big( \vec{p}\cdot \vec{x}_n \big)^a
 \,, \nonumber \\
&&\bar{I}_{\pi R} = \frac{\gamma_N^R-2}{16\,\pi^2} - \frac{1}{16\,\pi^2}\,\int_0^1\,\dd{z} \,\log\,\frac{F_{\pi R}(z)}{M_R^2} \sim Q
\,,\nonumber \\
&&X_{\pi R}^{(n_x)}[\,\vec{x}_n,\,\vec{p}\,] =  \frac{(-1)^{\lfloor (1-n_x) / 2 \rfloor}}{8\,\pi^2}\,\int_0^1 d\,z\,K_{n_x} \big[\, F_{\pi R}(z),\, \vert\vec{x}_n\vert \big] \,
 {\rm cs}_{n_x}\big( z\,\vec{ p }\cdot \vec{x}_n  \big)
 \,,\nonumber \\
&&F_{\pi R}(z) = z\,M_R^2 + (1-z)\,m_\pi^2- (1-z)\,z\,p^2\,, 
\label{def-bubble-basis}
\end{eqnarray}
where we use the short-hand notation ${\rm cs}_n(x)=\cos (x)$ for $n$ even, but ${\rm cs}_n(x)=\sin (x)$ for $n$ odd. The brackets $\lfloor x \rfloor := \operatorname{max}\{k\in\mathbb{Z}\,\vert\,k\leq z\}$ denote the floor function. With our expressions of Eq. (\ref{def-bubble-basis}) we separated the ultraviolet-convergent finite-box contributions from the ultraviolet-divergent infinite-volume contributions, $ \bar I_{L\pi}$ and $ \bar I_{\pi R}$. To ensure a consistent power counting for the two domains $m_\pi\sim \Delta$ and $m_\pi \ll \Delta$ the subtactions terms $\gamma^{L/R}_N$ are needed. In Ref. \cite{Hermsen:2024eth} a more detailed discussion is given, we only cite the results:
\begin{equation}
r = \frac{\Delta}{M}\,,\qquad
a = r\,(2+r)\,,\qquad
\gamma^{\Delta}_N = a\,\log\,\frac{1+a}{a}\,,\qquad
\gamma^{N}_{N} = 0\,.
\label{eqn:subtraction-bubble}
\end{equation}

We next discuss the bubble-loop integrals in terms of our in-box basis. Following our power-counting strategy the coefficients in front of the renormalized basis function are expanded according to Eq. (\ref{def-counting-rule}). The renormalized $\pi N$-contributions is then given by
\begin{eqnarray}
&&\bar{J}^A_{N\pi} + \bar{J}^A_{\pi N} = 
-\,m_\pi^2\,\Big(\renLQ[N \pi]{0}{0}{0} + \renQR[\pi N]{0}{0}{0}\Big) 
+ \renLQ[N \pi]{1}{0}{0}
- \renLQ[N \pi]{0}{0}{1} 
- \renQR[\pi N]{1}{0}{0}
+ \renQR[\pi N]{0}{0}{1} 
\notag \\ &&\hspace{3em}
+\,\frac{4}{3}\,\big(g_S - 2\,g_T\big)\,M_N\,m_\pi^2\,\Big(\renLQ[N \pi]{0}{0}{0} + \renQR[\pi N]{0}{0}{0}\Big) 
\notag \\ &&\hspace{3em}
-\,\frac{2}{3}\,\big(g_S+g_T\big)\,M_N\,\Big(\renLQ[N \pi]{0}{1}{0} + \renQR[\pi N]{0}{1}{0}\Big) 
-2\,\big(g_S+g_T\big)\,M_N\,\Big(\renSLQ[N \pi]{2} + \renSQR[\pi N]{2}\Big)
\notag \\ &&\hspace{3em}
+ \, 8\,g_F\,M_N\,\Big(
\renLQ[N \pi]{1}{0}{0} - \renLQ[N\pi]{0}{0}{1}
- \renQR[\pi N]{1}{0}{0}+\renQR[\pi N]{0}{0}{1} 
\Big) + \mathcal{O}(Q^4)\,.
\label{eqn:JA-NucleonBubble-contributions}
\end{eqnarray}

A few comments may be useful here. We want to emphasize, that the infinite-volume limit of the complete form factor is free of any kinematic singularities at all chiral orders.  In contrast, the finite-volume effects exhibit artificial kinematic singularities at $t = 4\,M_N^2$ and $t=0$, due to our projector in Eq. \eqref{ProjAxialcurrent}, which are not lifted by our reduction. While the first pole is far out of the momentum range of the chiral expansion and can therefore be expanded, the second is absorbed into the integrals
\begin{equation}\begin{aligned}[b]
\bar{S}_{\ldots}^{(2)} &= \frac{\renQR[\ldots]{0}{0}{2} - 2\,\renQR[\ldots]{1}{0}{1} + \renQR[\ldots]{2}{0}{0}}{t}\sim Q^3\,,
\label{eqn:defS2}
\end{aligned}\end{equation}
with $\cdots\in\{L\pi,\pi R\}$\,.
Having a non-vanishing upper index $\bar a=2$ or $a =2$ in Eq. (\ref{def-bubble-basis}), their finite-box values depend on either $\vec p$ or $ \vec {\bar p}$ and their infinite-volume parts are zero. By deriving a separate projector for $J_A(\vec{\bar{p}},\,\vec{p})$ at $t=0$ analog  to Eq. \eqref{ProjAxialcurrent}, we can proof that the linear combinations $S_{\ldots}^{(2)}$ are regular at $t=0$ with the limit 
\begin{eqnarray}
\lim_{t\to 0}S_{\ldots}^{(2)} =  -\,\frac{1}{3}\,\bar{I}_{\ldots}^{(0,1,0)} - \frac{1}{3\,M_N^2}\bar{I}_{\ldots}^{(1,0,1)}\,.
\label{res-t-limit-zero}
\end{eqnarray}
Finally, we would like to point out that the LEC $g_F$ in the last line of Eq. \eqref{eqn:JA-NucleonBubble-contributions}, does not appear in the infinite volume \cite{Lutz:2023xpi,Hermsen:2024eth,Schindler:2006it,Bernard:1998gv} and is therefore determined by the momentum dependency of the form factor in the finite volume.  In our example study, in Sec. \ref{sec:Numerical}, we set $g_F=0$ for convenience. 
We turn to the $\pi\Delta$-contribution by specifying their explicit expression as
\begin{eqnarray}
&&\bar{J}^A_{\Delta\pi} + \bar{J}^A_{\pi\Delta} = 
\frac{5}{9}\,f_A^{+}\,M_N\,\bigg\{
20\,\Big[-m_\pi^2\,\alpA{1}{2}{0} + 2\,M_N\,r\,\delta\,\alpA{1}{3}{0}\Big]\Big(\renLQ[\Delta\pi]{0}{0}{0} + \renQR[\pi\Delta]{0}{0}{0}\Big)
\notag \\ &&\hspace{6em}
+\,3\,r\,\alpA{1}{0}{1}\,\Big(\renLQ[\Delta\pi]{1}{0}{0} - \renLQ[\Delta\pi]{0}{0}{1} - \renQR[\pi\Delta]{1}{0}{0} + \renQR[\pi\Delta]{0}{0}{1}\Big)
\notag \\ &&\hspace{6em}
+\,\alpA{1}{0}{3}\,\Big(\renLQ[\Delta\pi]{0}{1}{0} + \renQR[\pi\Delta]{0}{1}{0}\Big)
+3\,\alpA{1}{0}{4}\,\Big(\renSLQ[\Delta\pi]{2} + \renSQR[\pi\Delta]{2}\Big)
\bigg\} 
\notag \\
\notag &&\hspace{3em}
-\,\frac{1}{9}\,f_A^{-}\,M_N\,\bigg\{
4\,\Big[m_\pi^2\,\alpA{2}{2}{0} - 2\,M_N\,r\,\delta\,\alpA{2}{3}{0}\Big]\,\Big(\renLQ[\Delta\pi]{0}{0}{0} + \renQR[\pi\Delta]{0}{0}{0}\Big)
\notag \\ &&\hspace{6em}
+\,3\,r\,\alpA{2}{0}{1}\Big(\renLQ[\Delta\pi]{1}{0}{0} - \renLQ[\Delta\pi]{0}{0}{1} - \renQR[\pi\Delta]{1}{0}{0} + \renQR[\pi\Delta]{0}{0}{1}\Big)
\notag \\ &&\hspace{6em}
+\,\alpA{2}{0}{3}\,\Big(\renLQ[\Delta\pi]{0}{1}{0} + \renQR[\pi\Delta]{0}{1}{0}\Big)
+3\,\alpA{2}{0}{4}\,\Big(\renSLQ[\Delta\pi]{2} + \renSQR[\pi\Delta]{2}\Big)
\bigg\}
\notag \\ &&\hspace{3em}
+\,\frac{8}{9}\,f_M\,M_N\,\bigg\{4\,r^2\,t\,\alpA{3}{1}{0}\,\Big(\renLQ[\Delta\pi]{0}{0}{0} + \renQR[\pi\Delta]{0}{0}{0} \Big)
\notag \\ &&\hspace{6em}
-\,3\,\alpA{3}{0}{1}\,\Big(\renLQ[\Delta\pi]{1}{0}{0} - \renLQ[\Delta\pi]{0}{0}{1} - \renQR[\pi\Delta]{1}{0}{0} + \renQR[\pi\Delta]{0}{0}{1}\Big)\bigg\} + \mathcal{O}\big(Q^4\big)\,.
\label{eqn:JA-DeltaBubble-contributions}
\end{eqnarray}
 In Eq. \eqref{eqn:JA-DeltaBubble-contributions} the factors $\alpha_{a \IS b}^{A,c}$ arise, which are detailed in Appendix \ref{app:alpFactors}. These coefficients are functions of $r$ only and normalized with $\alpha_{a \IS b}^{A,c}\to0$ at $r\to0$. In the infinite-volume limit we recover our previous bubble contribution \cite{Hermsen:2024eth,Lutz:2020dfi} by simply setting $\alpha_{a \IS b}^{A,c}=0$ if $c\neq 0$. Note that we suppressed the explicit dependency of our loop functions $\bar{J}^{A}_{\ldots}$ and basis integrals in Eqs. \eqref{eqn:JA-NucleonBubble-contributions}-\eqref{eqn:JA-DeltaBubble-contributions} out of notational simplicity, this is how we will continue from now on. \\

\par
Finally, we discuss the triangle terms. The basis functions take the form
\begin{eqnarray}
&&\bar I^{(\bar a,\, h , \,a)}_{L^m \pi R^n}(\vec{ \bar p}, \,\vec{p} \,) = \delta_{\bar a,0}\,\delta_{ h,0}\,\delta_{ a,0}\, \bar I_{ L^m \pi R^n}
\notag \\ && \hspace{8em}
+ \sum^{\vec n \neq 0}_{\vec{n}\in\mathbb{Z}^3} \,X_{L^m\pi R^n}^{(\bar{a} + 2\,h + a)}[\,\vec{\bar{p}},\,\vec{x}_n,\,\vec{p}\,]\,
 \big(\vec{\bar p}\cdot \vec{x}_n \big)^{\bar a}\,\big(\vec{x}_n^{\,2}\big)^h\, \big( \vec{p}\cdot \vec{x}_n \big)^a
 \,, \nonumber \\ 
&&\bar{I}_{L^m\! \pi R^n} = \frac{1}{16\,\pi^2} \int_0^1 d z_L \int_0^{1-z_L} d z_R\, \frac{z_L^m\,z_R^n}{ F_{L\pi R}(z_L,z_R) } - \frac{\gamma^{(m,n)}_{L\pi R}}{16\,\pi^2\, M^2} \sim Q^0
\,, \nonumber \\
&& X_{L^m \pi R^n}^{(n_x)}[\,\vec{\bar{p}},\,\vec{x}_n,\,\vec{p}\,] =
\frac{(-1)^{\lfloor (1-n_x ) / 2 \rfloor}}{16\,\pi^2}\,\int_0^1 \dd{z_L} \int_0^{1-z_L} \dd{z_R}
\nonumber \\ &&\hspace{8em}\times\,K_{-1 + n_x}\big[\, F_{L\pi R}(z_L,z_R),\, \vert\vec{x}_n\vert \big]\,
{\rm cs}_{n_x}\big( (z_L\,\vec{ \bar p} + z_R\,\vec{p}\,)\cdot \vec{x}_n \,\big)\,z_L^m\,z_R^n
\,, \nonumber\\
&&F_{L\pi R}(z_L,z_R) = m_\pi^2 
- z_L\,\big( \bar p^2 - M_L^2 + m_\pi^2  \big)  - z_R\,\big( p^2 - M_R^2 + m_\pi^2  \big)
\nonumber\\
&&\hspace{8em}
+\,z_L^2\,\bar{p}^2 +  z_L\,z_R\,(\bar{p}^2 +  p^2 - t) + z_R^2\,p^2\,, 
\label{def-triangle-basis}
\end{eqnarray}
where we find two options, either choose $m= 0$ or $n= 0$ always. The first term in Eq. (\ref{def-triangle-basis}) with $ \bar I_{ L^m\! \pi R^n} \sim Q^0 $, is ultraviolet finite always. Like in Ref. \cite{Hermsen:2024eth}, a finite subtraction is 
implemented to arrive at the scaling $\bar I_{ L^m\! \pi R^n} \sim Q^2 $ in the chiral domain $m_\pi \ll  M_\Delta-M_N $. We recall the required terms:
\begin{eqnarray}
&&\gamma^{(0,0)}_{\Delta\pi N}=\gamma^{(0,0)}_{N\pi\Delta}=
 \frac{1}{a}\,\log (1+ a) +\log \frac{ 1+a}{a} 
 \,,\nonumber\\
&&\gamma^{(1,0)}_{\Delta\pi N}=\gamma^{(0,1)}_{N\pi\Delta}= \frac{1+a}{3\,a}-\frac{1}{3\,a^2}\,
\log (1+ a) - \frac{a}{3}\,\log \frac{ 1+a}{a} 
\,,\nonumber\\
&&\gamma^{(2,0)}_{\Delta\pi N}=\gamma^{(0,2)}_{N\pi\Delta}= \frac{-2 + a +a^2-2\, a^3}{10\,a^2}+\frac{1}{5\,a^3}\,
\log (1+ a) + \frac{a^2}{5}\,\log \frac{ 1+a}{a} 
\,,\nonumber\\
&&\gamma^{(0,0)}_{\Delta \pi \Delta }=\log \frac{ 1+a}{a}
\,,\nonumber\\
&&\gamma^{(1,0)}_{\Delta \pi \Delta } = \gamma^{(0,1)}_{\Delta \pi \Delta }= \frac{1}{2}\,\Big( 1-a\,\log \frac{ 1+a}{a} \Big) 
\,,\nonumber\\
&&\gamma^{(2,0)}_{\Delta \pi \Delta }= 
\gamma^{(0,2)}_{\Delta \pi \Delta }= \frac{1}{6}\,\Big( 1-2\, a+2\,a^2\,\log \frac{ 1+a}{a} \Big)\,,
\end{eqnarray}
where the definition of $a$ is given in Eq. \eqref{eqn:subtraction-bubble}. Note that 
$\gamma^{(m,n)}_{N\pi N}= 0$ always. 
This allows us to specify the renormalized triangle parts to the axial-vector form factor. We begin by discussing the $N\pi N$-contribution, given by

\begin{eqnarray}
&&\bar{J}^A_{N\pi N} = \frac{1}{3}\,\bigg\{\,\renQ[\pi]{0}{0}{0} + m_\pi^2\,\Big(\renLQ[N\pi ]{0}{0}{0} + \renQR[\pi N]{0}{0}{0}\Big)
+2\,\Big(\renLQ[N\pi]{1}{0}{0} - \renLQ[N\pi]{0}{0}{1} 
\notag \\ &&\hspace{6em}
-\,\renQR[\pi N]{1}{0}{0} + \renQR[\pi N]{0}{0}{1} \Big)\bigg\}
\notag \\ &&\hspace{2em}
-\, \frac{4}{3}\,M_N^2\,\bigg\{\,m_\pi^2\,\renLQR[N-\pi-N]{0}{0}{0}{0}{0} 
+ \renLQR[N-\pi-N]{0}{0}{0}{1}{0}
+3\,\renSLQR[N-\pi-N]{0}{0}{2}
+\, 2\,\Big(\renLQR[N-\pi-N]{1}{0}{1}{0}{0} - \renLQR[N-\pi-N]{1}{0}{0}{0}{1}
\notag \\ &&\hspace{6em}
-\, \renLQR[N-\pi-N]{0}{1}{1}{0}{0} 
+ \renLQR[N-\pi-N]{0}{1}{0}{0}{1} \Big)
+ t\,\Big(\renLQR[N-\pi-N]{2}{0}{0}{0}{0} + \renLQR[N-\pi-N]{0}{2}{0}{0}{0}\Big)\bigg\} + \mathcal{O}(Q^4) \,.
\label{eqn:resJANpiN}
\end{eqnarray}
First, we would like to point out, that the infinite-volume limit of Eq. \eqref{eqn:resJANpiN} differs from our previous, already updated, result in Ref. \cite{Hermsen:2024eth}. This is due to the different handling of the artificial singularities, caused by the projector in Eq. \eqref{ProjAxialcurrent}, which we will discuss now in more detail. In Ref. \cite{Hermsen:2024eth} we used the Passarino-Veltman reduction scheme from Ref. \cite{Lutz:2020dfi} in the first step of the calculation. The $t=0$ singularities were then lifted by replacing $\bar{I}_{L^0\pi R^0}^{(0,\,0,\,0)}$ with a linear combination of $\bar{I}_{L^m\pi R^n}^{(0,\,0,\,0)}$, as derived in \cite{Isken:2023xfo}. As explained in more detail in Appendix \ref{app:redInBox}, these integrals are now an essential part in our new in-box reduction scheme and appear naturally in the final results. This automatically  leads to no singularities in the infinite-volume at $t=0$ or even $t=4\,M_N^2$ and therefore, no replacement rules are needed.  As explained at the end of Appendix \ref{app:redInBox} we ensure the reader, that the full expressions can be related, but  the chiral expansion may differ upon higher orders. Thus our novel in-box reduction scheme is already advantageous in the infinite-volume limit. The singularities appearing in the finite volume corrections are treated the same way as in the bubble. The $t=4\,M_N^2$ singularity is expanded, while the one at $t=0$ is encoded in the integral $\renSLQR[L-\pi-R]{m}{n}{2}$. The definition and limit follows from Eqs. \eqref{eqn:defS2} and  \eqref{res-t-limit-zero} by setting   $\cdots = L^m\pi R^n$. This concludes the discussion and we will continue by specifying the $N\pi\Delta$-contribution:
\begin{eqnarray}
&&\bar{J}^A_{\Delta \pi N} + \bar{J}^A_{N \pi \Delta}
 = 
-\,\frac{1}{9}\,f_S\bigg\{
\Big[r\,t\,\alpA{4}{1}{0} - 4\,m_\pi^2\,\alpA{4}{2}{0} + 8\,M_N\,\delta\,\alpA{4}{3}{0}\Big]\,\Big(\renLQ[\Delta\pi]{0}{0}{0} + \renQR[\pi\Delta]{0}{0}{0}\Big)
\notag \\ &&\hspace{8em}
+\,2\,\alpA{4}{0}{1}\,\Big(\renLQ[\Delta\pi]{1}{0}{0} - \renLQ[\Delta\pi]{0}{0}{1} - \renQR[\pi\Delta]{1}{0}{0} + \renQR[\pi\Delta]{0}{0}{1}\Big)
\notag \\ &&\hspace{8em}
-\,\frac{1}{2}\,\alpA{4}{0}{3}\,\Big(\renLQ[\Delta\pi]{0}{1}{0} + \renQR[\pi\Delta]{0}{1}{0}\Big)
-\frac{3}{2}\,\alpA{4}{0}{4}\,\Big(\renSLQ[\Delta\pi]{2} + \renSQR[\pi\Delta]{2}\Big)\bigg\}
\notag \\ &&\hspace{2em}
-\,\frac{2}{9}\,f_S\,M_N^2\bigg\{
8\,m_\pi^2\,\alpA{5}{2}{0}\,\Big(\renLQR[\Delta-\pi-N]{0}{0}{0}{0}{0}+\renLQR[N-\pi-\Delta]{0}{0}{0}{0}{0}\Big)
-\alpA{5}{0}{3}\,\Big(
\renLQR[\Delta-\pi-N]{0}{0}{0}{1}{0}
+\renLQR[N-\pi-\Delta]{0}{0}{0}{1}{0}
\Big)
\notag \\ &&\hspace{8em}
-\,3\,\alpA{5}{0}{4}\,\Big(
\renSLQR[\Delta-\pi-N]{0}{0}{2} + \renSLQR[N-\pi-\Delta]{0}{0}{2}
\Big)
-4 \,\alpA{6}{0}{1}\,\Big(
\renLQR[\Delta-\pi-N]{1}{0}{1}{0}{0}
-\renLQR[\Delta-\pi-N]{1}{0}{0}{0}{1}
\notag \\ &&\hspace{8em}
-\,\renLQR[N-\pi-\Delta]{0}{1}{1}{0}{0}
+\renLQR[N-\pi-\Delta]{0}{1}{0}{0}{1}
\Big)
-2\,t\,\alpA{7}{1}{0}\,\Big(
\renLQR[\Delta-\pi-N]{2}{0}{0}{0}{0}
+\renLQR[N-\pi-\Delta]{0}{2}{0}{0}{0}
\Big)
\bigg\}
\notag \\ &&\hspace{2em}
-\,\frac{4}{3}\,f_E\,M_N\,r\,\bigg\{
\frac{1}{3}\,t\,\alpA{8}{1}{0}\,\Big(\renLQ[\Delta\pi]{0}{0}{0} + \renQR[\pi\Delta]{0}{0}{0}\Big)
\notag \\ &&\hspace{8em}
-\,\frac{3}{2}\,\alpA{8}{0}{1}\,\Big(\renLQ[\Delta\pi]{1}{0}{0} - \renLQ[\Delta\pi]{0}{0}{1} - \renQR[\pi\Delta]{1}{0}{0} + \renQR[\pi\Delta]{0}{0}{1}\Big)
\bigg\}+ \mathcal{O}(Q^4)\,.
\label{eqn:resJADpiN}
\end{eqnarray}
By comparing Eq. \eqref{eqn:resJADpiN}  with our previous result in Ref. \cite{Hermsen:2024eth}, we identify two additional merits of our in-box reduction. First, we no longer have any $1/r$ terms contributing in the kinematic functions of the basis elements in the infinite-volume limit. Second, integrals of the type $I_{N\pi}^{(\bar{a},\,h,\,a)}$ do not appear any more, which is due to our choice of triangle basis integrals. We close this section by providing the $\Delta\pi\Delta$-contribution:
\begin{eqnarray}
&&\bar{J}^A_{\Delta \pi \Delta}
 = 
\frac{1}{27}\,\bigg\{
\Big[r\,t\,\alpA{9}{1}{0} + 32\,r\,m_\pi^2\,\alpA{9}{2}{0} - 20\,M_N\,\delta\,\alpA{9}{3}{0}\Big]\,\Big(\renLQ[\Delta\pi]{0}{0}{0} + \renQR[\pi\Delta]{0}{0}{0}\Big)
\notag \\ &&\hspace{6em}
+\,2\,\alpA{9}{0}{1}\,\Big(
\renLQ[\Delta\pi]{1}{0}{0}
-\renLQ[\Delta\pi]{0}{0}{1}
-\renQR[\pi\Delta]{1}{0}{0}
+\renQR[\pi\Delta]{0}{0}{1}
\Big)
\notag \\ &&\hspace{6em}
-\,\alpA{9}{0}{3}\,\Big(\renLQ[\Delta\pi]{0}{1}{0} + \renQR[\pi\Delta]{0}{1}{0}\Big)
-3\,\alpA{9}{0}{4}\,\Big(\renSLQ[\Delta\pi]{2} + \renSQR[\pi\Delta]{2} \Big)
\bigg\}
\notag \\ &&\hspace{2em}
-\,\frac{4}{27}\,M_N^2\bigg\{
\Big[4\,r^2\,t\,\alpA{10}{1}{0} + 10\,m_\pi^2\,\alpA{10}{2}{0} - 20\,M_N\,r\,\delta\,\alpA{10}{3}{0}\Big]\,\renLQR[\Delta-\pi-\Delta]{0}{0}{0}{0}{0}
\notag \\ &&\hspace{6em}
+\,\alpA{10}{0}{3}\,\renLQR[\Delta-\pi-\Delta]{0}{0}{0}{1}{0}
+3\,\alpA{10}{0}{4}\,\renSLQR[\Delta-\pi-\Delta]{0}{0}{2}
+\,6\,r\,t\,\alpA{11}{1}{0}\,\Big(\renLQR[\Delta-\pi-\Delta]{1}{0}{0}{0}{0}+\renLQR[\Delta-\pi-\Delta]{0}{1}{0}{0}{0}\Big)
\notag \\ &&\hspace{6em}
+\,2\,\alpA{11}{0}{1}\,\Big(
\renLQR[\Delta-\pi-\Delta]{1}{0}{1}{0}{0}
-\renLQR[\Delta-\pi-\Delta]{1}{0}{0}{0}{1}
-\renLQR[\Delta-\pi-\Delta]{0}{1}{1}{0}{0}
+\renLQR[\Delta-\pi-\Delta]{0}{1}{0}{0}{1}
\Big)
\notag \\ &&\hspace{6em}
+\,t\,\alpA{12}{1}{0}\,\Big(\renLQR[\Delta-\pi-\Delta]{2}{0}{0}{0}{0}+\renLQR[\Delta-\pi-\Delta]{0}{2}{0}{0}{0}\Big)
\bigg\}+ \mathcal{O}(Q^4)\,.
\label{eqn:JA-triangle-contributions}
\end{eqnarray}

\clearpage 

\section{Numerical results}
\label{sec:Numerical}

\begin{figure*}[b]
    \centering
    \includegraphics[width=\textwidth]{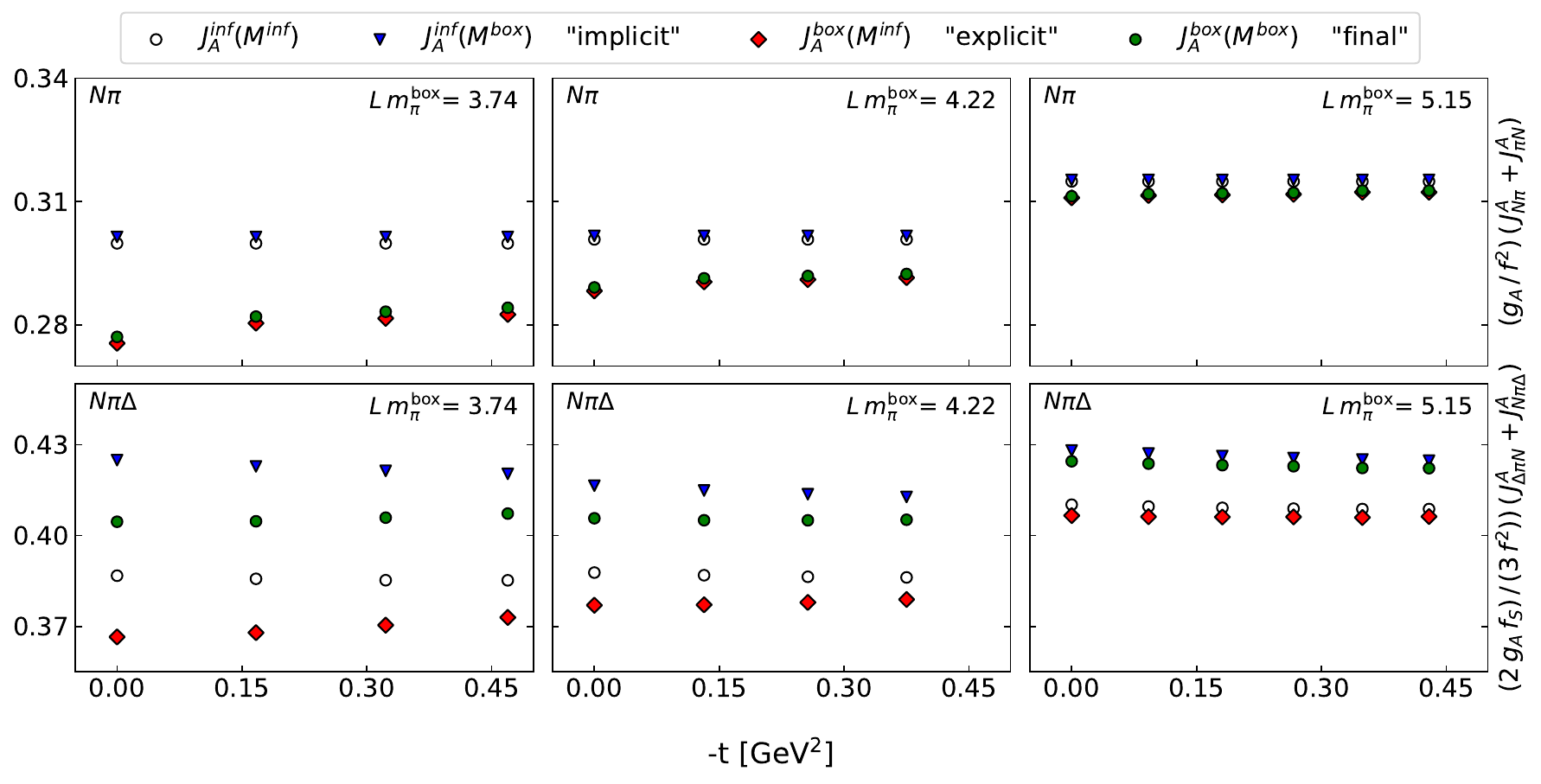}
	\caption{The nucleon bubble (top) and nucleon-pion-$\Delta$ loop (bottom) corrections to the axial-vector form factor as a function of $t = (\bar p - p)^2)$. The lattice length $L$ increases from the left to the right. The white circles correspond to the infinite-volume case, with no finite-box corrections in the masses or in the form factor. The blue triangles contain only the finite-box corrections of the hadron masses. These ``implicit'' finite-box effects were already considered in Ref. \cite{Lutz:2020dfi}. The red diamonds contain only the ``explicit'' finite-box corrections to the form factor, by using the infinite-volume hadron masses. The green dots give the full results, taking both types of finite-box corrections into account.}
	\label{fig:nucleonBubAndNpiDTri}
\end{figure*} 

We present selected numerical results for three lattice ensembles, with the pion mass in all cases around 250 MeV, and the box size increasing from 2.98 fm via 3.36 fm to 4.04 fm. In Fig.~\ref{fig:nucleonBubAndNpiDTri} we plot the two largest loop corrections to the form factor: the $N\pi$ bubble and the $N\pi\Delta$ triangle from Eq. \eqref{res-GA}. The LECs are taken from Ref. \cite{Hermsen:2024eth} and listed in Table \ref{tab:listLECs}.
The hadron masses used are given in Table \ref{tab:dataForNumeric}. From Fig.~\ref{fig:nucleonBubAndNpiDTri}, we observe that for the $N\pi$ bubble the implicit finite-box effects of the nucleon and pion mass are small, since the blue triangles and the white circles as well as the green circles and the red diamonds are close to each other. Therefore, the full finite-box corrections are dominated by the explicit contributions to the form factor. These corrections are also the only source of the $t$-dependence, since there is no explicit dependence to the chiral order considered. The non-linear behavior stems from the $\vec{p}$-dependence of our basis elements  (\ref{def-bubble-basis}) which is projected onto $t$. 
\\
\par

\begin{table*}[t]
    \begin{tabular}{lp{5pt}ccp{10pt}ccp{10pt}cc}
    \toprule
        && \multicolumn{2}{c}{$L\,m_\pi=3.74$} & &\multicolumn{2}{c}{$L\,m_\pi=4.22$} &&
        \multicolumn{2}{c}{$L\,m_\pi=5.15$} 
        \\\cmidrule{3-4}\cmidrule{6-7}\cmidrule{9-10} 
        && \,in-box & $\infty$  && \,in-box & $\infty$ && \,in-box & $\infty$\\\midrule
        $m_\pi\,$[\si{\mega\electronvolt}] && 248 & 247 && 248 & 248 && 252 & 252 \\
        $M_N\,$[\si{\giga\electronvolt}] &&1.036&1.044&&1.037&1.044&&1.042&1.048 \\
        $M_\Delta\,$[\si{\giga\electronvolt}] &&1.374&1.342&&1.367&1.342&&1.363&1.347 \\
        $L\,$[fm] &&2.98& - & & 3.36 & - & & 4.04 &- \\
        Fitted in \cite{Lutz:2020dfi} &&\multicolumn{2}{c}{no} && \multicolumn{2}{c}{yes} &&\multicolumn{2}{c}{yes} \\ 
        lattice ensemble &&\multicolumn{2}{c}{ETMC \cite{EuropeanTwistedMass:2008pab,Alexandrou:2010hf}} && \multicolumn{2}{c}{CLS O7\,\cite{Capitani:2017qpc,Capitani:2015sba}} &&\multicolumn{2}{c}{CLS B6\,\cite{Capitani:2017qpc,Capitani:2015sba}} \\  
        \\\bottomrule\\
    \end{tabular}
    \caption{
     Hadron masses and lattice lengths of the ensembles used in our numerical analysis of the axial-vector form factor. For each ensemble we give in in-box masses, taken from Ref. \cite{Lutz:2020dfi}. The pion mass in the infinite volume was calculated. The corresponding nucleon and $\Delta$-isobar masses are taken from Ref. \cite{Lutz:2020dfi}.
    The second to last line indicates if the corresponding form-factor data were used in the global fit of Ref. \cite{Lutz:2020dfi}. The last line referrers to the corresponding lattice ensemble.}
    \label{tab:dataForNumeric}
\end{table*}

In the smallest box with $L\,m_\pi^{\text{box}}=3.74$ we observe finite-box corrections of about 8-10\%, while in the largest box with $L\,m_\pi^{\text{box}}=5.15$ they decrease to about 1-2\%. This is in line with the  $\exp(-m_\pi\,L)$ behavior of the implicit effects and the small difference between the in-box and infinite-volume mass of the nucleon. The full finite-box corrections of the $N\pi\Delta$ triangles show a more complicated behavior. Due to the large effects resulting from the use of the in-box $\Delta$-mass, the full finite-box correction of the form factor now result from the explicit and implicit effects. This can be seen in the plot as the red diamonds approach the white circles, while the green circles approach the blue triangles. By increasing the lattice size the implicit effects start to dominate the full corrections, due to the $\exp(-m_\pi\,L)$ suppression of the explicit effects. Furthermore, we observe a negative contribution of the explicit effects, but a positive one of the implicit effects, which results in an overall positive correction.
\\
\par
While we observe a relative correction of about 5\% in the smallest lattice box, which interestingly is smaller than the relative correction of the $N\pi$ bubbles, we find a relative correction of the same order in the largest box with $L\,m_\pi^{\text{box}} = 5.15$. This is in contrast to the $N\pi$ bubble, where the relative correction drops significantly with increasing $L\,m_\pi^{\text{box}}$. The dependence on $t$ is more complex, since we have now an explicit $t$-dependence in the kinematical function of the form factors. Since the relative corrections of the explicit effects decrease faster than the ones of the implicit ones, the implicit ones become more dominant for higher momentum transfers. Altogether, the full finite-box corrections demonstrate a non-linear behavior again. 
\\
\par
These two examples of one-loop contributions clearly indicate that the in-box mass of the $\Delta$-isobar plays an important role in the determination of the in-box axial-vector form factor. It even needs to be considered at $L\,m_\pi^{\text{box}}=5.15$, which is in line with our previous work \cite{Lutz:2020dfi}, where only the implicit effects were considered. Furthermore, we conclude that the use of on-shell in-box masses and the full finite-box corrections of the axial-vector form factor are crucial for precise results.

\clearpage
\section{Summary}
\label{sec:Summary}
In this work we developed a novel framework to compute form factors in a finite box. We derived a minimal basis set of one-loop functions that can be used to express finite-box effects of LQCD observables in terms of on-shell hadron masses.
The use of such masses implies a particular summation scheme that leaves the strict path of conventional perturbative chiral expansions. It involves isobar mass parameters with finite-box effects that are not always exponentially suppressed by large volumes.  
While such a scheme appears to accelerate the chiral expansion significantly, the exponential suppression of volume effects, as it is seen for form factors in strict perturbative approaches, is not always manifest here.
Specific results for the axial-vector form factor of the nucleon were derived at the one-loop level. 
A systematic study of such finite-box effects is needed to constrain the LECs of the chiral Lagrangian. 
\\
\par
We exemplified the impact of finite-box effects with flavor-SU(2) ensembles for which baryon masses and form factor data are available. The relative role of implicit versus explicit finite-box effects was investigated. In most cases we confirm our previous working hypothesis \cite{Lutz:2020dfi,Hermsen:2024eth} that finite-box effects are dominated by the implicit effects that are caused by using the finite-box masses of the $\Delta$-isobar in the computation of the form factor. The explicit contributions lead to a striking momentum dependence that calls for dedicated analyses of LQCD data. 
\\
\par
For precise results on the axial-vector form factor it would be best to use ensembles at physical strange-quark masses such that the more precise flavor-SU(2) chiral extrapolation framework is applicable. Since the $\Delta$-isobar degree of freedom plays an important role in that form factor  accurate lattice results on the nucleon and $\Delta$-isobar masses in not too large boxes are most useful.  
 To connect LQCD data at large quark masses with the conditions met in the laboratory our 
results can play an important role. The data set required 
in our approach is more sustainable than the very expensive data sets with pion masses below 200 MeV and large box sizes.

\newpage
\appendix
\section{The basis of loop functions in the finite volume}
\label{app:redInBox}
In this Section we will derive a tensor decomposition in the finite volume and present a reduction into a minimal set of basis integrals for generic tadpole, bubble, and triangle integrals. A calculation of the finite-volume corrections of these one-loop tensor integrals can be found in Refs. \cite{Lozano:2020qcg,Liang:2022tcj}, but the results are only given in an over-complete set of basis integrals. These integrals relate to our intermediate basis functions given in \eqref{def-kernel}. If applied to the axial-vector form factor, one would end up with a plethora of integrals, which is numerically not feasible. Furthermore, both these works treated the infinite-volume and the finite-volume corrections separately in terms of tensor decomposition, calculation, and final basis elements, although it is possible to do it simultaneously, as it was proven in Ref. \cite{Lutz:2014oxa} for the baryon masses. \\

\par
The target function of our tensor decomposition is given by
\begin{eqnarray}
&&J_{\bar{k}}^{(\bar{a},\,h,\,a)}[\bar{p},\,m^2,\,\vec{v},\,\vec{x}_n,\,p]
\notag \\ &&\qquad 
= (-1)^{\bar{k}-1}\,i\,\Gamma(\bar{k})\,\ensuremath{\mu^{4-d}\,  \int\frac{\dd[d]{l}}{(2\pi)^{d}}} \frac{e^{i\,(\vec{l}+\vec{v})\,\cdot\,\vec{x}_n}\,\big(\bar{p}\cdot l\big)^{\bar{a}}\,\big(l^2\big)^h\,\big(l\cdot p\big)^{a}}{\big(l^2-m^2 + i\,\varepsilon\big)^{\bar{k}}}\,.
\label{app:eqn:targetFkt}
\end{eqnarray}
Note that this is a function of $\vec{x}_n$. The sum over $\mathbb{Z}^3$ is not relevant for the tensor decomposition and is therefore dropped. It is straightforward to derive the decomposition in the infinite-volume limit with $\vec{n}=\vec{0}$  using the usual tensor ansatz due to Lorentz invariance. The decomposition in the finite-volume corrections, with $\vec{n}\neq \vec{0}$ is more challenging, since broken rotational and translational symmetries. To achieve such a decomposition, we need to split any $d$-dimensional object into their energy-like and three-momentum-like components. To do this properly, we apply the so-called split-dimensional regularization \cite{Leibbrandt:1996np}. Here the dimension $d$ is split into  $d = d_t+d_s$ with $d_t= 1 + \delta_t$ and $d_s = 3+\delta_s$. All result need to be finite in the limit $\delta_t,\,\delta_s\to 0$. The $d$-dimensional vectors are then split according to 
\begin{equation}
l\cdot p= l_0\cdot p_0 - \vec{l}\cdot\vec{p}\,,
\end{equation} 
where the $0$ in index denotes a $d_t$-dimensional vector and the vector arrow a $d_s$-dimensional vector. Note that for the vector $x$ per definition $x_0=0$, since only the space is discretized, where the time is left continuous. After splitting all momenta in our target function in \eqref{app:eqn:targetFkt} for $\vec{x}_n\neq 0$, we first need to solve the energy-like integration. It is straightforward to apply dimensional regularization to the integral in question and find the identity
\begin{equation}\begin{aligned}[b]
&\int\frac{\dd[d_t]{l_0}}{(2\pi)^{d_t}}\,\frac{l_0^{2\,h}}{\big(l_0^2-\vec{l}^2 - m^2 + i\,\varepsilon\big)^{\bar{k}}} 
\\ &\qquad 
=  \frac{1}{2^{d_t}\,\Gamma(\bar{k})\,\Gamma(d_t/2)\,\pi^{d_t/2-1}} \frac{\Gamma(h+d_t/2)}{\Gamma\big(h-\bar{k}+(d_t+2)/2\big)}\,\frac{(-1)^{d_t}i}{\big[\vec{l}^2 + m^2 \big]^{\bar{k} - h  - d_t/2}}\,.
\end{aligned}\end{equation}
The remaining $d_s$ momentum terms can then be determined with a tensor-decomposition operating in the $d_s$ dimensional subspace,
\begin{eqnarray}
\label{app:tensorDecompMomFVE}
&&\int\frac{\dd[d_s]{l}}{(2\,\pi)^{d_s}}\frac{e^{i\,(\vec{l}\,\cdot\,\vec{x}_n)}\,\vec{l}^{j_1}\,\cdots\,\vec{l}^{j_k}}{\big[\vec{l}^2+m^2\big]^{\bar{k} - 1/2 }}
= 
\sum_{v=0/1}^k\,\frac{\Gamma(\bar{k} - (k+v)/2 - 1/2 )}{2^{\nicefrac{(k+v)}{2}}\,\Gamma(\bar{k}-1/2)}\,T_v^{j_1\ldots j_k}
\notag \\ &&\hspace{16em}\times\,
\int\frac{\dd[d_s]{l}}{(2\,\pi)^{d_s}}\,\frac{i^v\,e^{i\,(\vec{l}\,\cdot\,\vec{x}_n)}}{\big[\vec{l}^2+m^2\big]^{\bar{k}-(k+v)/2 - 1/2}}\,.
\end{eqnarray}
Here we introduced the fully-symmetric tensor $T_v^{j_1j_2\ldots j_m}$ which contains $v$ open $\vec{x}_n^{(j_i)}$. As examples we give
\begin{equation}\begin{gathered}[b]
T_0^{j_1j_2} = \delta^{j_1j_2}\,,\qquad T_2^{j_1j_2} = \vec{x}_n^{(j_1)}\,\vec{x}_n^{(j_2)}\,,\\
T_1^{j_1j_2j_3} = \vec{x}_n^{(j_1)}\,\delta^{j_2j_3}+\vec{x}_n^{(j_2)}\,\delta^{j_1j_3}+\vec{x}_n^{(j_3)}\,\delta^{j_1j_2}\,,\qquad
T_3^{j_1j_2j_3} = \vec{x}_n^{(j_1)}\,\vec{x}_n^{(j_2)}\,\vec{x}_n^{(j_3)}\,.
\end{gathered}\end{equation}
The relation in Eq. \eqref{app:tensorDecompMomFVE} can be proven by induction, using a suitable recursive decomposition of the tensor $T_v^{j_1j_2\ldots j_k}$ and differentiating with respect to an arbitrary momentum $\vec{k}_i$. This proof is motivated by Ref. \cite{Greil:2011aa}, where a related identity was determined for one and two open loop momenta only. 
\\ \par
Combining the tensor decomposition of the infinite-volume limit and from the finite-volume corrections we then establish, after some tedious algebra, the basis identity of our finite volume decomposition scheme:
\begin{equation}\begin{aligned}[b]
&J_{\bar{k}}^{(\bar{a},\,h,\,a)}[\bar{p},\,m^2,\,\vec{v},\,\vec{x}_n,\,p]
\\&\hspace{1em}
 = \mu^{4-d}\,(-1)^{\bar{k}-1}\,i\,\Gamma(\bar{k})\,\int\frac{\dd[d]{l}}{(2\,\pi)^{d}}\,\frac{e^{i(\vec{l}+\vec{v})\cdot\vec{x}_n}\,\big(\bar{p}\cdot  l\big)^{\bar{a}}\,\big(l^2\big)^h\,\big(l\cdot p\big)^{a}}{\big[l^2-m^2+i\,\varepsilon\big]^{\bar{k}}}\\
&\hspace{1em}
 = \sum_{\bar{b},\bar{h},b,k}^{\infty}\,C^{(\bar{a},\,h,\,a)}_{\bar{b}\,\bar{h}\,b,k}
\,\big(\bar{p}\cdot p\big)^k\,\big(\vec{x}_n^2\big)^{h-\bar{h}}\,
\bar{p}^{\bar{b}-k}\,\big(\vec{\bar{p}}\cdot\vec{x}_n\big)^{\bar{a}-\bar{b}}
 \\ &\hspace{6em}\times
\,V_{\bar{k},\bar{h}+\frac{\bar{b}+b}{2}}^{(\bar{a}-\bar{b} + 2\,(h-\bar{h}) + a-b)}[m^2,\,\vec{x}_n,\,\vec{v}\,]\,\big(\vec{p}\cdot\vec{x}_n\big)^{a-b}\,p^{b-k}
\,\,,\\\\
&C^{(\bar{a},\,h,\,a)}_{\bar{b}\,\bar{h}\,b,k} = 
\begin{dcases}
&0\,, \text{ if } \quad
	\bar{b}-k\text{ odd }\quad
 \vert\vert\quad
 b-k\text{ odd }
 \quad\vert\vert \quad
\bar{b}+b \text{ odd }\,,\\ 
&k!\binom{\bar{a}}{\bar{b}}\,\binom{\bar{b}}{k}\,(\bar{b}-k-1)!!\,\binom{a}{b}\,\binom{b}{k}\,(b-k-1)!!\,\binom{h}{\bar{h}}
\\ &
\times\,(-1)^{h+\bar{h}}\, \prod_{n=0}^{\bar{h}-1}\big(d+2(n + \bar{a} + h-\bar{h} +a)-\bar{b}-b\big)
\,,\text{ else }
\end{dcases}
\end{aligned}\end{equation}
with $\bar b - h$, $b- h$ and $\bar b + b$ even always and e.g. $5 !! = 1\times  3\times 5 $. The generalize basis elements are defined as
\begin{equation}\begin{aligned}[b]
&V_{\bar{k},k}^{(n_x)}[m^2,\,\vec{x}_n,\,\vec{v}\,] = I_{\bar{k}}^{(k)}[m^2]\,\delta_{n_x,0}\,\delta_{\vec{n},\vec{0}} + X_{\bar{k},k}^{(n_x)}[m^2,\,\vec{x}_n,\,\vec{v}\,]\,(1-\delta_{\vec{n},\vec{0}})\,,\\[.75em]
&I^{(k)}_{\bar{k}}[m^2] = (-1)^{\bar{k}-1}\,\Gamma(\bar{k})\Bigg[\prod_{n=0}^{k-1}\frac{1}{d+2\,n}\Bigg]\,\int\frac{\dd[d]{l}}{(2\,\pi)^{d}}\frac{\mu^{4-d}\,i\,\big(l^2\big)^{k}}{\big(l^2-m^2+i\,\varepsilon\big)^{\bar{k}}}\,,\\[.75em]
&X^{(n_x)}_{\bar{k},k}[m^2,\,\vec{x}_n,\,\vec{v}\,]
= (-1)^{n_x+k+(1+3\,d_t)/2}\,\frac{\Gamma\big(\bar{k}-k-n_x - d_t/2\big)}{2^{k+n_x+d_t}\,\pi^{d_t/2}}
\\ &\hspace{10em}  \times 
\,\int\frac{\dd[d_s]{l}}{(2\,\pi)^{d_s}}\,\frac{\mu^{4-d_t-d_s}\,i^{n_x}\,e^{i\,(\vec{l}+\vec{v})\cdot\vec{x}_n}}{\big[\vec{l}^{2}+m^2\big]^{\bar{k}-k-n_x - d_t/2}}
\\ 
&\phantom{X^{(n_x)}_{\bar{k},k}[m^2,\,\vec{x}_n,\,\vec{v}\,]}\,
= (-1)^{n_x+k}\,\frac{i^{n_x}\,e^{i\,\vec{v}\cdot\vec{x}_n}}{2^{\bar{k}-k+1}\,\pi^{2}}
K_{2-\bar{k}+n_x+k}[m^2, \vert \vec{x}_n\vert\,]\,.
\label{app:eqn-defGenBasis}
\end{aligned}\end{equation}
We want to comment the last line of Eq. \eqref{app:eqn-defGenBasis}. Here we applied Eq. (A.13) from Ref. \cite{Hasenfratz:1989pk} and Eq. (3.471.9) from Ref. \cite{gradshteyn2014table} and took the limits $d_t\to 1$ and $d_s\to 3$. While our tensor decomposition and the following reduction scheme holds for any value of $m^2$, the rewrite into the modified Bessel function is only valid for 
$\Im(m^2)\neq 0$  and  $\Re(m^2)>0$. In the following, we consider two types of heavy fields $L,R$ and one light field $Q$. In the main part of this work, we set $L,R\in\{N,\Delta\}$ and $Q=\pi$. Deduced from the generic basis, we introduce a further set of functions, which we will be calling kernels
\begin{eqnarray}
&&\kernelQ{k}{n_x}\big[\,\vec{x}_n\,\big] = V_{1,k}^{(n_x)}\big[m_Q^2,\,\vec{x}_n,\,\vec{0}\,\big]\,, \notag \\
&&
\kernelLQ{n}{k}{n_x}\big[\,\vec{\bar{p}},\,\vec{x}_n\,\big]
= \int_0^1\,\dd{z_L}\,z_L^n\,V^{(n_x)}_{2,k}\big[F_{LQ}(z_L),\,\vec{x}_n,\,z_L\,\vec{\bar{p}}\,\big]
\,,\notag \\
&&\kernelQR{n}{k}{n_x}\big[\,\vec{x}_n,\,\vec{p}\,\big]  = \int_0^1\,\dd{z_R}\,z_R^{n}\,V^{(n_x)}_{2,k}\big[F_{QR}(z_R),\,\vec{x}_n,\,z_R\,\vec{p}\,\big]  \,,\notag \\
&&\kernelLQR{m}{n}{k}{n_x}\big[\,\vec{\bar{p}} ,\,\vec{x}_n,\, \vec{p} \, \big]
\notag\\ &&\qquad 
 = 
\int_0^1\,\dd{z_L}\,\int_0^{1-z_L}\,\dd{z_R}\,z_L^{m}\,z_R^{n}\,V_{3,k}^{(n_x)}\big[F_{LQR}(z_L,\,z_R),\,\vec{x}_n,\,z_L\,\vec{\bar{p}} + z_R\,\vec{p}\,\big]\,,\notag \\
&&  F_{L Q}(z_L) = z_L\,M_L^2 + (1-z_L)\,m_Q^2- (1-z_L)\,z_L\,\bar p^2\,,       
\notag \\
&& F_{Q R}(z_R) = z_R\,M_R^2 + (1-z_R)\,m_Q^2 - (1-z_R)\,z_R\,p^2\,,
\notag \\
&&F_{L Q R}(z_L,z_R) = m_Q^2 
- z_L\,\big( \bar p^2 - M_L^2 + m_Q^2  \big)  - z_R\,\big( p^2 - M_R^2 + m_Q^2  \big)
\notag \\
&&\hspace{9.5em}
+\,z_L^2\,\bar{p}^2 + z_L\,z_R\,(\bar{p}^2 +  p^2 - t) + z_R^2\,p^2\,.
\label{def-kernel}
\end{eqnarray}
Note that $F_{L Q}(z_L) = F_{L Q R}(z_L,0)$ and $F_{Q R}(z_R) = F_{L Q R}(0,z_R)$ hold. These elements are subject of our reduction scheme and are defining our unrenormalized in-box basis elements
\begin{eqnarray}
&&I_{Q}^{(\bar{a},\,h,\,a)} = \sum\limits_{\vec{n}\in\mathbb{Z}^3}\,\big(\vec{\bar{p}}\cdot\vec{x}_n\big)^{\bar{a}}\,\big(\vec{x}_n^2\big)^{h}\,\big(\vec{p}\cdot\vec{x}_n\big)^{a}\,\kernelQ{0}{\bar{a}+2\,h+a}[\,\vec{x}_n\,]\,,\notag\\
&&I_{L Q}^{(\bar{a},\,h,\,a)} = \sum\limits_{\vec{n}\in\mathbb{Z}^3}\,\big(\vec{\bar{p}}\cdot\vec{x}_n\big)^{\bar{a}}\,\big(\vec{x}_n^2\big)^{h}\,\big(\vec{p}\cdot\vec{x}_n\big)^{a}\,\kernelLQ{0}{0}{\bar{a}+2\,h+a}[\,\vec{\bar{p}},\,\vec{x}_n\,]\,,\notag\\
&&I_{QR}^{(\bar{a},\,h,\,a)}= \sum\limits_{\vec{n}\in\mathbb{Z}^3}\,\big(\vec{\bar{p}}\cdot\vec{x}_n\big)^{\bar{a}}\,\big(\vec{x}_n^2\big)^{h}\,\big(\vec{p}\cdot\vec{x}_n\big)^{a}\,\kernelQR{0}{0}{\bar{a}+2\,h+a}[\,\vec{x}_n,\,\vec{p}\,]  \,,\notag\\
&&I_{L^mQR^n}^{(\bar{a},\,h,\,a)} = \sum\limits_{\vec{n}\in\mathbb{Z}^3}\,\big(\vec{\bar{p}}\cdot\vec{x}_n\big)^{\bar{a}}\,\big(\vec{x}_n^2\big)^{h}\,\big(\vec{p}\cdot\vec{x}_n\big)^{a}\,\kernelLQR{m}{n}{0}{n_x}\big[\,\vec{\bar{p}} ,\,\vec{x}_n, \, \vec{p}^{\,2} \, \big]\,.
\label{def-intermidate-feynman}
\end{eqnarray}
Their renormalized forms for $Q\to\pi$ are detailed in the Eqs. (\ref{def-tadpole-basis}), (\ref{def-bubble-basis}), and (\ref{def-triangle-basis}),  respectively. 
Additionally, we want to clarify that we used the phrase \say{in-box} instead of \say{finite volume}, due to simplicity. In the first step of our reduction scheme we need to express the over-complete basis given in Eq. \eqref{def-overcomplete-basis} into our kernel functions. We decompose
\begin{equation}
J^{(\bar{a} \IS h \IS a)}_{\ldots} = \sum\limits_{\vec{n}\in\mathbb{Z}^3}\,J^{(\bar{a} \IS h \IS a)}_{\ldots}[\vec{x}_n],
\end{equation}
and find
\begin{eqnarray}
&&J_Q^{(\bar{a}\IS h\IS a)}[\vec{x}_n]
=\hspace{-0.3em}\sum_{\bar{b},\bar{h},b,k=0}^{\infty}\,C^{(\bar{a},\,h,\,a)}_{\bar{b}\,\bar{h}\,b,k}
\,\big(\bar{p}\cdot p\big)^k\,\big(\vec{x}_n^2\big)^{h-\bar{h}}\,
\bar{p}^{\bar{b}-k}\,\big(\vec{\bar{p}}\cdot\vec{x}_n\big)^{\bar{a}-\bar{b}}
\notag\\ &&\hspace{3em}\times\, 
\kernelQ{\bar{h}+\frac{\bar{b}+b}{2}}{\bar{a}-\bar{b} + 2\,(h-\bar{h}) + a-b}
\,\big(\vec{p}\cdot\vec{x}_n\big)^{a-b}\, p^{b-k}\,,\notag \\
&&J_{LQ}^{(\bar{a}\IS h\IS a)}[\vec{x}_n]=\hspace{-0.3em}\sum_{\vert\mathbf{\bar{c}}\vert,\,\vert\mathbf{\bar{e}}\vert,\,\vert\mathbf{c}\vert}^{\bar{a},h,a}\,\sum_{\bar{b},b,{\bar{h}},k=0}^{\infty}\,\binom{\bar{a}}{\mathbf{\bar{c}}}\,\binom{h}{\mathbf{\bar{e}}}\,\binom{a}{\mathbf{c}}\,C^{(\bar{c}_1 + \bar{e}_2,\, \bar{e}_1,\,c_1)}_{\bar{b}\,{\bar{h}}\,b,k}\,2^{\bar{e}_2}
\notag\\ &&\hspace{3em} \times\,
\big(\bar{p}\cdot p\big)^{k+\bar{c}_2}\,\big(\vec{x}_n^2\big)^{\bar{e}_1-\bar{h}}
\,\bar{p}^{\bar{b}-k+2\,\bar{c}_2+2\,\bar{e}_3}
\,\big(\vec{\bar{p}}\cdot\vec{x}_n\big)^{\bar{c}_1+\bar{e}_2-\bar{b}}
\notag\\ &&\hspace{3em}
\times\,\kernelLQ{\bar{c}_2+\bar{e}_2  +2\,\bar{e}_3+c_2 }{\bar{h}+\frac{\bar{b}+b}{2}}{\bar{c}_1+\bar{e}_2-\bar{b}+2\,(\bar{e}_1-\bar{h})+c_1-b}\,\big(\vec{p}\cdot\vec{x}_n\big)^{ c_1-b}\, p^{b-k}\,,\notag\\
&&J_{QR}^{(\bar{a}\IS h\IS a)}[\vec{x}_n]=\hspace{-0.3em}\sum_{\vert\mathbf{\bar{c}}\vert,\,\vert\mathbf{\bar{e}}\vert,\,\vert\mathbf{c}\vert}^{\bar{a},h,a}\,\sum_{\bar{b},b,{\bar{h}},k=0}^{\infty}\,\binom{\bar{a}}{\mathbf{\bar{c}}}\,\binom{h}{\mathbf{\bar{e}}}\,\binom{a}{\mathbf{c}}\,C^{(\bar{c}_1,\, \bar{e}_1,\,\bar{e}_2+c_1)}_{\bar{b}\,{\bar{h}}\,b,k}\,2^{\bar{e}_2}
\notag \\ &&\hspace{3em} \times\,
\big(\bar{p}\cdot p\big)^{k+\bar{c}_2}\,\big(\vec{x}_n^2\big)^{\bar{e}_1-\bar{h}}
\,\bar{p}^{\bar{b}-k}\,\big(\vec{\bar{p}}\cdot\vec{x}_n\big)^{\bar{c}_1-\bar{b}}
\notag \\ &&\hspace{3em} \times\,
\kernelQR{\bar{c}_2+\bar{e}_2  +2\,\bar{e}_3+c_2 }{\bar{h}+\frac{\bar{b}+b}{2}}{\bar{c}_1-\bar{b}+2\,(\bar{e}_1-\bar{h})+\bar{e}_2 + c_1-b}\,\big(\vec{p}\cdot\vec{x}_n\big)^{\bar{e}_2 + c_1-b}\, p^{b-k+2\,\bar{e}_3+2\,c_2}\,,\notag \\
&&J_{LQR}^{(\bar{a}\IS h\IS a )}[\vec{x}_n] =\hspace{-0.3em} 
\sum_{\vert\mathbf{\bar{c}}\vert,\,\vert\mathbf{\bar{e}}\vert,\,\vert\mathbf{c}\vert}^{\bar{a},h,a}\,\sum_{\bar{b},\bar{h},b,k=0}^{\infty}\,\binom{\bar{a}}{\mathbf{\bar{c}}}\,\binom{h}{\mathbf{\bar{e}}}\,\binom{a}{\mathbf{c}}\,C^{(\bar{c}_1+\bar{e}_2,\,\bar{e}_1,\,\bar{e}_3+c_1)}_{\bar{b}\,h\,b,k}\,2^{\bar{e}_2+\bar{e}_3+\bar{e}_5}
\notag \\ &&\hspace{3em} \times 
\,\big(\bar{p}\cdot  p\big)^{k + \bar{c}_3+\bar{e}_5+c_2}\,\big(\vec{x}_n^2\big)^{\bar{e}_1-\bar{h}}
\,\bar{p}^{\bar{b} -k + 2\,\bar{c}_2+2\,\bar{e}_4 }
\,\big(\vec{\bar{p}}\cdot\vec{x}_n\big)^{\bar{c}_1 + \bar{e}_2 - \bar{b}}
\notag \\ &&\hspace{3em} \times 
\,\kernelLQR{\bar{c}_2+\bar{e}_2+2\,\bar{e}_4+\bar{e}_5 + c_2}{\bar{c}_3+\bar{e}_3+\bar{e}_5+2\,\bar{e}_6+c_3}{\bar{h}+\frac{\bar{b}+b}{2}}{\bar{c}_1+\bar{e}_2-\bar{b} + 2\,(\bar{e}_1-\bar{h}) + \bar{e}_3+c_1-b}
\notag \\ &&\hspace{3em} \times \,
\big(\vec{p}\cdot\vec{x}_n\big)^{\bar{e}_3 + c_1 - b}
\, p^{b - k + 2\,\bar{e}_6+2\,c_3}\,,
 \label{res-overcomplete-basis}
\end{eqnarray}
where we use the multi-indices
\begin{equation}
\mathbf{\bar{c}} = (\bar{c}_1,\,\ldots,\,\bar{c}_i)\,,\quad
\mathbf{\bar{e}} = (\bar{e}_1,\,\ldots,\,\bar{e}_j)\,,\quad
\mathbf{c} = (c_1,\,\ldots,\,c_i)\,,
\end{equation}
with $i=2,j=3$ if the integral is a bubble and $i=3,\, j=6$ if it is a triangle, respectively. Furthermore we used the short-hand notations
\begin{equation}
\sum_{\vert\mathbf{\bar{c}}\vert,\,\vert\mathbf{\bar{e}}\vert,\,\vert\mathbf{c}\vert}^{\bar{a},h,a} = 
\sum_{\vert\mathbf{\bar{c}}\vert=\bar{a}}
\sum_{\vert\mathbf{\bar{e}}\vert=h}
\sum_{\vert\mathbf{c}\vert=a}\,, \qquad
\sum_{\vert\alpha\vert=k} = \sum_{\substack{\alpha_1+\ldots + \alpha_n = k \\ \alpha_1,\,\ldots\,,\alpha_n\geq 0} }\,,\qquad 
\binom{k}{\alpha} = \binom{k}{\alpha_1,\,\ldots,\,\alpha_n}\,.
\end{equation}
So far we expressed our initial most generic over-complete basis in terms of a further set of over-complete basis functions, the kernels defined in Eq. \eqref{def-kernel}. The merit of this step lies in the observation, that set of kernels can be reduced to our final set by means of particular recursion relations similar to the ones advocated previously in Ref. \cite{Isken:2023xfo} for the infinite-volume case and properly generalized to the finite-volume case in our current work. Before considering the different types of integrals we offer a reduction for the generic integral $V^{(n_x)}_{\bar{k},k}[m^2,\,\vec{x}_n,\,\vec{v}\,]$  in the index $k$ 
\begin{equation}\begin{aligned}[b]
V^{(n_x)}_{\bar{k},k} = \frac{1}{d  + 2\,(k-\bar{k} + n_x)}\,\Big\{ m^2\,V^{(n_x)}_{\bar{k},k-1} + \vec{x}_n^2\,V^{(n_x+2)}_{\bar{k},k-1}\Big\}\,.
\label{app:eqn:genericRedk}
\end{aligned}\end{equation}
This equation can be derived by using dimensional regularization, the tensor decomposition, and algebraic reductions. It is then straightforward to find the reduction of the tadpole kernel
\begin{equation}\begin{aligned}[b]
\kernelQ{k}{n_x} = \frac{1}{d  + 2\,(k - 1  + n_x)}\,\Big\{ m_Q^2\,\kernelQ{k-1}{n_x} + \vec{x}_n^2\,\kernelQ{k-1}{n_x+2}\Big\}\,.
\label{app:eqn:iterationTadpole}
\end{aligned}\end{equation}

We continue with the reduction for the bubbles kernels. In this case we can reduce the indices $k$ and $n$.  Adopting Eq. \eqref{app:eqn:genericRedk} for both bubbles types separately we each find a relation which decreases $k$ but also increases $n$.  Using partially integration individually onto both bubble kernels we obtain a second equation in each case, which decreases $n$ but increases $k$.  The final sets of reductions are determined by a specific combination for each bubble and given by 
\begin{eqnarray}
&&\kernelLQ{n}{k}{n_x} = 
\frac{1}{d+2\,(k-1+n+n_x)}\,\bigg\{m_Q^2\,\kernelLQ{n}{k-1}{n_x}-\bar{p}^2\,\kernelLQ{n+2}{k-1}{n_x} 
-\kernelL{k-1}{n_x} 
\notag \\ &&\hspace{16em} 
-\,2\,\big(\vec{\bar{p}}\cdot\vec{x}_n\big)\,\kernelLQ{n+1}{k-1}{n_x+1}  + \vec{x}_n^2\,\kernelLQ{n}{k-1}{n_x+2}\bigg\}\,,\notag \\
&&\kernelLQ{n}{k}{n_x} =
\frac{d+2\,(k-2+n+n_x)}{d-3+n+2\,(k+n_x)}\,\frac{\bar{p}^2 - M_L^2 + m_Q^2}{2\,\bar{p}^2}\,\kernelLQ{n-1}{k}{n_x}
\notag \\ && \hspace{3em}
-\,\frac{d+2\,(k-1+n_x)}{d-3+n+2\,(k+n_x)}\,\bigg\{
\frac{\vec{\bar{p}}\cdot\vec{x}_n}{\bar{p}^2}\,\kernelLQ{n-1}{k}{n_x+1}
-\frac{\delta_{n,1}}{2\,\bar{p}^2}\,\kernelQ{k}{n_x}
+\frac{1}{2\,\bar{p}^2}\,\kernelL{k}{n_x}\bigg\}
\notag \\ &&\hspace{3em} 
+\,\frac{1-n}{d-3+n+2\,(k+n_x)}\,\bigg\{\frac{m_Q^2}{\bar{p}^2}\,\kernelLQ{n-2}{k}{n_x} - \frac{\vec{x}_n^2}{\bar{p}^2}\,\kernelLQ{n-2}{k}{n_x+2}\bigg\}\,,\notag \\[1em ]
&&\kernelQR{n}{k}{n_x} = \frac{1}{d+2\,(k-1+n+n_x)}\,\bigg\{
m_Q^2\,\kernelQR{n}{k-1}{n_x}
- p^2\,\kernelQR{n+2}{k-1}{n_x}
-\kernelR{k-1}{n_x}
\notag \\ &&\hspace{16em} 
-\, 2\,\big(\vec{p}\cdot\vec{x}_n\big)\,\kernelQR{n+1}{k-1}{n_x+1}
+ \vec{x}_n^2\,\kernelQR{n}{k-1}{n_x+2}\bigg\}\,,\notag \\
&&\kernelQR{n}{k}{n_x} =
\frac{d + 2\,(k - 2 +n+n_x)}{d - 3 + n + 2\,(k+n_x)}\,\frac{ p^2 - M_R^2 + m_Q^2}{2\, p^2}\,\kernelQR{n-1}{k}{n_x}
\notag \\ && \hspace{3em}
-\,\frac{d + 2\,(k-1+n_x)}{d - 3 + n + 2\,(k+n_x)}\,\bigg\{\frac{\vec{p}\cdot\vec{x}_n}{ p^2}\,\kernelQR{n-1}{k}{n_x+1}
-\frac{\delta_{n,1}}{2\, p^2}\,\kernelQ{k}{n_x}
+\frac{1}{2\, p^2}\,\kernelR{k}{n_x}\bigg\}
\notag \\ && \hspace{3em}
+\,
\frac{1-n}{d - 3 + n + 2\,(k+n_x)}\,\bigg\{\frac{m_Q^2}{ p^2}\,\kernelQR{n-2}{k}{n_x}
- \frac{\vec{x}_n^2}{ p^2}\,\kernelQR{n-2}{k}{n_x+2}\bigg\}\,.
\end{eqnarray}
Note that we introduced the heavy-tadpole kernels
\begin{eqnarray}
&&\kernelQ[L]{k}{n_x} = V_{1,k}^{(n_x)}[M_L^2,\,\vec{x}_n,\,\vec{\bar{p}}\,]\,, \qquad
\kernelQ[R]{k}{n_x} = V_{1,k}^{(n_x)}[M_R^2,\,\vec{x}_n,\,\vec{p}\,]\,.
\end{eqnarray}

Their index $k$ can be reduced analog to Eq. \eqref{app:eqn:iterationTadpole} by replacing $m_Q\to M_{L/R}$. Since the finite-volume effects are suppressed by $e^{-M_B\,L}$, with $M_B\in\{M_N,M_\Delta\}$ these terms are systematically dropped. The only remaining ultraviolet contribution encoded in 
$I_{L}^{(0,0,0)}$ and $I^{(0,0,0)}_{R}$ is then also dropped upon renormalization, as it was done in our previous works \cite{Hermsen:2024eth,Lutz:2020dfi,Isken:2023xfo,Lutz:2014oxa,Semke:2005sn}.\\

\par
We now move on to the reduction of the triangle kernels. In this case we reduce the index $k$ to zero and either $m$ or $n$, demanding no artificial kinematic singularities emerging during the reduction, like in the infinite-volume limit \cite{Isken:2023xfo}. The initial three relations are given by adapting Eq. \eqref{app:eqn:genericRedk} to the triangle case and partially integrating over both Feynman parameters separately. See Eq. (B10) from Ref. \cite{Isken:2023xfo} for the relations deduced from partial integration in the infinite-volume limit.  The final reductions are subsequently obtained by a specific combination of the three aforementioned relations. The index  $k$ is reduced via
\begin{eqnarray}
&&\kernelLQR{m}{n}{k}{n_x}
=  \frac{1}{d  + 2\,(k- 1 + m + n + n_x)}\,\bigg\{m_Q^2\,\kernelLQR{m}{n}{k-1}{n_x}
\notag \\ &&\hspace{1em}
-\,\bar{p}^2\,\kernelLQR{m+2}{n}{k-1}{n_x}
-2\,\big(\bar{p}\cdot p\big)\,\kernelLQR{m+1}{n+1}{k-1}{n_x}
-  p^2\,\kernelLQR{m}{n+2}{k-1}{n_x}
\notag\\&&\hspace{1em}
-\,2\,\big(\vec{\bar{p}}\cdot\vec{x}_n\big)\,\kernelLQR{m+1}{n}{k-1}{n_x+1}
-2\,\big(\vec{p}\cdot\vec{x}_n\big)\,\kernelLQR{m}{n+1}{k-1}{n_x+1}
+\vec{x}_n^2\,\kernelLQR{m}{n}{k-1}{n_x+2}
\notag\\&&\hspace{1em}
-\,\kernelSigma{m+1}{n}{k-1}{n_x}
-\kernelSigma{m}{n+1}{k-1}{n_x}\bigg\}\,,
\end{eqnarray}
which is a generalization of  Eq. (B13) from \cite{Isken:2023xfo}. There the infinite-volume limit was considered only. In their notation it holds
\begin{equation}
    I_{L^{m}\pi R^{n},\,k}^{(0, \, 0, \,  0)} = \Bigg[\prod_{j=0}^{k-1}\frac{1}{d + 2\,j }\Bigg]\,I_k(m,n)\,,
\end{equation}
with $I_k(m,n)$ defined in Eq. (B2) of \cite{Isken:2023xfo}. As mentioned before, we can either reduce the index $m$ or $n$ in 
\kernelLQR{m}{n}{k}{n_x} to zero. The reduction of the index $m$ is given by
\begin{eqnarray}
&&\kernelLQR{m}{n}{k}{n_x}
= \frac{d  + 2\,(k-2 + m + n + n_x)}{d - 3 + m  + 2\,(k + n + n_x)}\,\frac{\bar{p}^2 - M_L^2+m_Q^2}{2\,\bar{p}^2}\,\kernelLQR{m-1}{n}{k}{n_x}
\notag\\ &&\quad
-\,\frac{ d  + 2\,(k-1  + n + n_x)}{d - 3 + m  + 2\,(k + n + n_x)}
\bigg\{\frac{\bar{p}\cdot p}{\bar{p}^2}\,\kernelLQR{m-1}{n+1}{k}{n_x} 
-\frac{\delta_{m,1}}{2\,\bar{p}^2}\,\kernelQR{n}{k}{n_x}
\notag\\ &&\hspace{2em} 
+\,\frac{\vec{\bar{p}}\cdot\vec{x}_n}{\bar{p}^2}\,\kernelLQR{m-1}{n}{k}{n_x+1}
+\frac{1}{2\,\bar{p}^2}\,
\kernelSigma{m-1}{n}{k}{n_x}
\bigg\}
\notag\\ &&\quad
+\,\frac{1-m}{d - 3 + m  + 2\,(k + n + n_x)}\,\bigg\{\frac{m_Q^2}{\bar{p}^2}\,\kernelLQR{m-2}{n}{k}{n_x}
- \frac{p^2}{\bar{p}^2}\,\kernelLQR{m-2}{n+2}{k}{n_x}
\notag\\&&\qquad
-\,\frac{2\,\big(\vec{p}\cdot\vec{x}_n\big)}{\bar{p}^2}\,\kernelLQR{m-2}{n+1}{k}{n_x+1}
+\frac{\vec{x}_n^2}{\bar{p}^2}\,\kernelLQR{m-2}{n}{k}{n_x+2}
-\,\frac{1}{\bar{p}^2}\,
\kernelSigma{m-2}{n+1}{k}{n_x}
 \bigg\}\,.
\end{eqnarray}
We introduced here the heavy-bubble kernel $\kernelSigma{m}{n}{k}{n_x}$. Formally we can reduce the index $k$ to zero and then renormalize all remaining heavy bubbles to zero in the infinite volume \cite{Isken:2023xfo}. Since the finite-volume corrections are  exponentially suppressed by the baryon masses, they can be dropped. 
Finally, we offer the reduction of the  triangle kernel function in the index $n$
\begin{eqnarray}
&&\kernelLQR{m}{n}{k}{n_x}
= 
\frac{d  + 2\,(k - 2 + m + n  + n_x)}{d  - 3 + n + 2\,(k + m  + n_x)}\,\frac{p^2-M_R^2+m_Q^2}{2\,p^2}\,\kernelLQR{m}{n-1}{k}{n_x}
\notag \\&&\quad
-\,\frac{d  + 2\,(k - 1 + m  + n_x) }{d  - 3 + n + 2\,(k + m  + n_x)}
\bigg\{
\frac{\bar{p}\cdot p}{p^2}\,\kernelLQR{m+1}{n-1}{k}{n_x}
-\frac{\delta_{n,1}}{2\,p^2}\,\kernelLQ{m}{k}{n_x}
\notag \\ &&\hspace{2em}
+\,\frac{\vec{p}\cdot\vec{x}_n}{p^2}\,\kernelLQR{m}{n-1}{k}{n_x+1}
+\,\frac{1}{2\,p^2}\,\kernelSigma{m}{n-1}{k}{n_x}
\bigg\}
\notag \\ &&\quad 
+\,\frac{1-n}{d  - 3 + n + 2\,(k + m  + n_x)}\,\bigg\{\frac{m_Q^2}{p^2}\,\kernelLQR{m}{n-2}{k}{n_x} 
-\frac{\bar{p}^2}{p^2}\,\kernelLQR{m+2}{n-2}{k}{n_x}
\notag \\&&\hspace{2em}
-\,\frac{2\,\big(\vec{\bar{p}}\cdot\vec{x}_n\big)}{p^2}\,\kernelLQR{m+1}{n-2}{k}{n_x+1}
+\frac{\vec{x}_n^2}{p^2}\,\kernelLQR{m}{n-2}{k}{n_x+2}
-\,\frac{1}{p^2}\,\kernelSigma{m+1}{n-2}{k}{n_x}
\bigg\}\,.
\end{eqnarray}

As explained in section \eqref{sec:Basis}, the chiral expansion from our previous works \cite{Hermsen:2024eth,Lutz:2020dfi} differ from the infinite-volume limit of our updated results.  
To be able to relate our results, we offer a reduction from
$I_{L^{0}\pi R^{n},\,k}^{(\bar{a}, \, h, \,  a)}$ to $I_{L^{0}\pi R^{n-1},\,k}^{(\bar{a}, \, h, \,  a)}$ and $I_{L^{0}\pi R^{n-2},\,k}^{(\bar{a}, \, h, \,  a)}$. We find 
\begin{eqnarray}
&&\big((\bar{p}\cdot p)^2 - \bar{p}^2\, p^2 \big)\,I_{L^{0}\pi R^{n},\,k}^{(\bar{a}, \, h, \,  a)}
= 
\frac{d - 5 + 2\,(k+n)+ 2\,(\bar{a}+2\,{h}+ a)}{2\,\big(d + n +2\,(k-2) + 2\,(\bar{a}+2\,{h}+ a)\big)}
\nonumber \\ &&\hspace{3em}\times\,
\Big( (\bar{p}\cdot p)\,(\bar{p}^2-M_L^2 + m_\pi^2)
-\,\bar{p}^2\,(p^2-M_R^2 + m_\pi^2)\Big)\,I_{L^{0}\pi R^{n-1},\,k}^{(\bar{a}, \, h, \,  a)}
\nonumber \\ &&\hspace{.5em}
+\,\frac{d-3 + 2\,k + 2\,(\bar{a}+2\,{h}+a)}{d + n + 2\,(k-2) + 2\,(\bar{a}+2\,{h}+ a)}\,
\nonumber \\ &&\hspace{3em}\times\,
\bigg\{
\bar{p}^2\,I_{L^{0}\pi R^{n-1},\,k}^{(\bar{a}, \, h, \,  a+1)}
-\,\big(\bar{p}\cdot p\big)\,I_{L^{0}\pi R^{n-1},\,k}^{(\bar{a}+1, \, h, \,  a)}
+\,\frac{\bar{p}^2}{2}\,\Big(\Sigma_{m-1,\,0,\,k}^{(\bar{a},\,h,\, a)} - \delta_{1,n}\,I_{L^0Q,\,k}^{(\bar{a},\, h\, a)}\Big)\bigg\}
\nonumber\\ &&\hspace{.5em}
+\,\frac{n-1}{d + n +2\,(k-2) + 2\,(\bar{a}+2\,{h}+ a)}\,\bigg\{
\Big(m_\pi^2\,\bar{p}^2 - \frac{1}{4}\,\big(\bar{p}^2-M_L^2 + m_\pi^2\big)^2\Big)\,I_{L^{0}\pi R^{n-2},\,k}^{(\bar{a}, \, h, \,  a)}
\nonumber\\ &&\hspace{1.5em}
-\,\bar{p}^2\,I_{L^{0}\pi R^{n-2},\,k}^{(\bar{a}, \, h+1, \,  a)}
+I_{L^{0}\pi R^{n-2},\,k}^{(\bar{a}+2, \, h, \,  a)}
+\frac{1}{4}\,(\bar{p}^2-M_L^2 + m_\pi^2)\,\Big(\Sigma_{0,\,n-2,\,k}^{(\bar{a},\,h,\, a)} - I_{\pi R^{n-2},\,k}^{(\bar{a}, \, h, \,  a)}\Big)
\nonumber\\ &&\hspace{1.5em}
+\,\frac{1}{2}\,\Big(\Sigma_{0,\,n-2,\,k}^{(\bar{a}+1,\,h,\, a)}-I_{\pi R^{n-2},\,k}^{(\bar{a}+1, \, h, \,  a)} \Big)
-\frac{1}{2}\,\bar{p}^2\,\Sigma_{1,\,n-2,\,k}^{(\bar{a},\,h, \, a)}\bigg\}
\nonumber\\ &&\hspace{1.5em}
+\frac{1}{2}\,(\bar{p}\cdot p)\,\bigg\{ I_{\pi R^{n-1},\,k}^{(\bar{a}, \, h, \,  a)} - \Sigma_{0,\,n-1,\,k}^{(\bar{a},\,h,\, a)}\bigg\}\,,
\end{eqnarray}
which is a generalization of Eqs. (34) and (B12) from Ref. \cite{Isken:2023xfo}\,. The corresponding reduction for the index $m$ is given by 
\begin{eqnarray}
&&\big((\bar{p}\cdot p)^2-\bar{p}^2\, p^2\big)\,I_{L^{m}\pi R^{0},\,k}^{(\bar{a}, \, h, \,  a)}
= 
\frac{d - 5 + 2\,(k+m) + 2\,(\bar{a}+2\,{h}+ a)}{2\,\big(d + m +2\,(k-2) + 2\,(\bar{a}+2\,{h}+ a)\big)}
\nonumber \\ &&\hspace{3em}\times\,
\Big( (\bar{p}\cdot p)\,(p^2-M_R^2 + m_\pi^2) 
- \,p^2\,(\bar{p}^2-M_L^2 + m_\pi^2)\Big)
\,I_{L^{m-1}\pi R^{0},\,k}^{(\bar{a}, \, h, \,  a)}
\nonumber\\ &&\hspace{.5em}
+\,\frac{d-3+2\,k + 2\,(\bar{a}+2\,{h}+a)}{d + m + 2\,(k-2) + 2\,(\bar{a}+2\,{h}+ a) }\,
\nonumber \\ &&\hspace{3em}\times\,
\bigg\{p^2\,I_{L^{m-1}\pi R^{0},\,k}^{(\bar{a}+1, \, h, \,  a)}
-\,\big(\bar{p}\cdot p\big)\,I_{L^{m-1}\pi R^{0},\,k}^{(\bar{a}, \, h, \,  a+1)}
+\,\frac{p^2}{2}\,\Big(\Sigma_{m-1,\,0,\,k}^{(\bar{a},\,h,\, a)} - \delta_{1,m}\,I_{QR^0,\,k}^{(\bar{a},\, h\, a)}\Big)\bigg\}
\nonumber\\ &&\hspace{.5em} 
+\,\frac{m-1}{d + m + 2\,(k-2) + 2\,(\bar{a}+2\,{h}+ a)}\,\bigg\{
\Big(m_\pi^2\,p^2 - \frac{1}{4}\,(p^2-M_R^2+m_\pi^2\big)^2\Big)\,I_{L^{m-2}\pi R^{0},\,k}^{(\bar{a}, \, h, \,  a)}
\nonumber\\ &&\hspace{1.5em} 
-\,p^2\,I_{L^{m-2}\pi R^{0},\,k}^{(\bar{a}, \, h+1, \,  a)}
+I_{L^{m-2}\pi R^{0},\,k}^{(\bar{a}, \, h, \,  a+2)}
+\frac{1}{4}\,\big(p^2-M_R^2 + m_\pi^2\big)\,\Big(\Sigma_{m-2,\,0,\,k}^{(\bar{a},\,h, \,a)} - I_{L^{m-2}\pi,\,k}^{(\bar{a}, \, h, \,  a)}\Big)
\nonumber\\ &&\hspace{1.5em} 
+\,\frac{1}{2}\,\Big(\Sigma_{m-2,\,0,\,k}^{(\bar{a},\,h, \, a+1)}-I_{L^{m-2}\pi,\,k}^{(\bar{a}, \, h, \,  a+1)}\Big)-\frac{p^2}{2}\,\Sigma_{m-2,\,1,\,k}^{(\bar{a},\,h, \, a)}\bigg\}
\nonumber\\ &&\hspace{1.5em}
+\frac{1}{2}\,(\bar{p}\cdot p)\,\bigg\{I_{L^{m-1}\pi,\,k}^{(\bar{a},\,h,\,a)}-\Sigma_{m-1,\,0,\,k}^{(\bar{a},\,h, \,a)}\bigg\}\,.
\end{eqnarray}

\clearpage
\section{List of $\alpha$ factors}
\label{app:alpFactors}
We give the $\alpha_{a,b}^{A,c}$ factors used in Eqs.
\eqref{eqn:JA-DeltaBubble-contributions}, \eqref{eqn:resJADpiN} and \eqref{eqn:JA-triangle-contributions}:
\begin{longtable}[b]{CC}
\alpha_{1,2}^{A,0} =  \frac{(2+r)^2\,(5+r)\,(4+4\,r+3\,r^2)}{80\,(1+r)^2}\,,& \hspace{-6.5em}
\alpha_{1,3}^{A,0} =  \frac{(2+r)^3\,(20 + 36\,r + 29\,r^2 + 5\,r^3)}{160\,(1+r)^3}\,,\\[1.25em]
\alpha_{1,0}^{A,1}  = \frac{(2+r)^2\,(1+2\,r)}{4\,(1+r)^2}\,,& \hspace{-6.5em}
\alpha_{1,0}^{A,3}  = \frac{(2+r)\,(1+r+r^2)}{2\,(1+r)^2} \\[1.25em]
\alpha_{1,0}^{A,4}  =  \frac{2+3\,r + 3\,r^2 + r^3}{2\,(1+r)^2}\,,  \\[1.25em]
\alpha_{2,2}^{A,0} =  \frac{(1-r)\,(2+r)^2\,(4+4\,r+3\,r^2)}{16\,(1+r)^2}\,,&\hspace{-6.5em}
\alpha_{2,3}^{A,0} =  \frac{(2+r)^3\,(4-5\,r^2-5\,r^3)}{32\,(1+r)^3}\,,  \\[1.25em]
\alpha_{2,0}^{A,1}  =  \frac{(2+r)^2}{4\,(1+r)^2}\,,&\hspace{-6.5em}
\alpha_{2,0}^{A,3} = \alpha_{2,0}^{A,4} =  \frac{(2+r)\,(1+r+r^2)}{2\,(1+r)^2}\,,  \\[1.25em]
\alpha_{3,1}^{A,0} =  \frac{(2+r)^4}{16\,(1+r)^2}\,,&\hspace{-6.5em}
\alpha_{3,0}^{A,1}  =  \frac{(2+r)^2\,(1+r+r^2)}{4\,(1+r)^2}\,,  \\[1.25em]
\alpha_{4,1}^{A,0} =  \frac{(2+r)^2\,(5-3\,r-5\,r^2)}{20\,(1+r)^2}\,,&\hspace{-6.5em}  \alpha_{4,2}^{A,0} = \frac{16 + 64\,r + 64\,r^2+22\,r^3+3\,r^4}{16\,(1+r)^2}\\[1.25em]
\alpha_{4,3}^{A,0} = \frac{(2+r)^4\,(1+2\,r+2\,r^2)}{16\,(1+r)^3}\,,&\hspace{-6.5em} \alpha_{4,0}^{A,1} = \frac{(2+r)\,(20 + 67\,r + 35\,r^2)}{40\,(1+r)^2}\,,  \\[1.25em]
\alpha_{4,0}^{A,3} = \alpha_{4,0}^{A,4} = \frac{1 - r - r^2}{(1+r)^2}\,,&\hspace{-6.5em} \alpha_{5,2}^{A,0} = \alpha_{5,0}^{A,3} = \alpha_{5,0}^{A,4} = \frac{2+r}{2}\,,  \\[1.25em]
\alpha_{6,0}^{A,1} = \frac{2+r}{2}\,,&\hspace{-6.5em}
\alpha_{7,1}^{A,0} = \frac{2+r}{2}\,, \\[1.25em]
\alpha_{8,1}^{A,0} = \frac{(2+r)^2\,(1+3\,r+r^2)}{4\,(1+r)}\,, &\hspace{-6.5em} \alpha_{8,0}^{A,1} = \frac{6+9\,r+5\,r^2+r^3}{6\,(1+r)^2} \,,\\[1.25em] 
\multicolumn{2}{l}{$\displaystyle\alpha_{9,1}^{A,0} = \frac{(2+r)^3\,(10+141\,r+236\,r^2+101\,r^3+15\,r^4)}{80\,(1+r)^4}$\,,}  \\[1.25em]
\multicolumn{2}{l}{$\displaystyle\alpha_{9,2}^{A,0} = \frac{(2+r)\,(64 + 132\,r + 113\,r^2+49\,r^3 + 8\,r^4)}{128\,(1+r)^4}$\,,}  \\[1.25em]
\multicolumn{2}{l}{$\displaystyle \alpha_{9,3}^{A,0} = \frac{(2+r)^2\,(20+80\,r+211\,r^2+293\,r^3+218\,r^4+84\,r^5+12\,r^6)}{80\,(1+r)^5}$\,,}  \\[1.25em]
\multicolumn{2}{l}{$\displaystyle\alpha_{9,0}^{A,1} =\frac{(2+r)^2\,(10+36\,r + 43\,r^2 + 18\,r^3 + 7\,r^4)}{40\,(1+r)^4}$\,,}  \\[1.25em]
\alpha_{9,0}^{A,3} = \alpha_{9,0}^{A,4} = \frac{2+7\,r + 9\,r^2 + 7\,r^3 + 2\,r^4}{2\,(1+r)^4}\,,& \\[1.25em]
\alpha_{10,1}^{A,0} = \frac{(2+r)^4\,(40 + 195\,r + 227\,r^2 + 97\,r^3 + 12\,r^4)}{640\,(1+r)^4}\,,&\\[1.25em]
\alpha_{10,2}^{A,0} = \frac{(2 + r)^2\, (20 + 40\,r + 39\,r^2 + 19\,r^3 + 2\,r^4)}{80\,(1+r)^2}\,,& \\[1.25em]
\alpha_{10,3}^{A,0} =\frac{(2 + r)^3\, (20 + 60\,r + 69\,r^2 + 31\,r^3 + 3\,r^4)}{160\,(1+r)^3}\,,& \\[1.25em]
\alpha_{10,0}^{A,3} = \alpha_{10,0}^{A,4} = \frac{(2+r)^2\,(2 + 2\,r - r^2)}{8\,(1+r)^2}\,, &  \\[1.25em]
\alpha_{11,1}^{A,0} = \frac{(2+r)^3\,(20+51\,r+58\,r^2+28\,r^3+2\,r^4)}{160\,(1+r)^4}\,,& \\[1.25em]
\alpha_{11,0}^{A,1} = \frac{(2+r)^2\,(2+2\,r-r^2)}{8\,(1+r)^2}\,, & \hspace{-6.5em} \alpha_{12,1}^{A,0} = \frac{(2+r)^2\,(2+2\,r-r^2)}{8\,(1+r)^2} \,.\\[1.25em]

\end{longtable}

\clearpage
\typeout{}
\bibliography{literature} 

@article{Luscher:1986pf,
  title = {Volume Dependence of the Energy Spectrum in Massive Quantum Field Theories},
  author = {Lüscher, M.},
  date = {1986},
  journaltitle = {Communications in Mathematical Physics},
  shortjournal = {Commun. Math. Phys.},
  volume = {105},
  number = {2},
  pages = {153--188},
  issn = {1432-0916},
  doi = {10.1007/BF01211097},
  url = {https://doi.org/10.1007/BF01211097},
  langid = {english}
}

@article{Hermsen:2025,
    author = "Hermsen, Felix and Lutz, Matthias F. M. and Timmermans, Rob G. E.",
    title = {In preperation},
}

@article{Doring:2011ip,
  title = {Dynamical Coupled-Channel Approaches on a Momentum Lattice},
  author = {Döring, M. and Haidenbauer, J. and Meißner, U.-G. and Rusetsky, A.},
  date = {2011},
  journaltitle = {The European Physical Journal A},
  shortjournal = {Eur. Phys. J. A},
  volume = {47},
  number = {12},
  eprint = {1108.0676},
  eprinttype = {arXiv},
  eprintclass = {hep-lat},
  pages = {163},
  issn = {1434-6001, 1434-601X},
  doi = {10.1140/epja/i2011-11163-7},
  url = {http://arxiv.org/abs/1108.0676},
  urldate = {2021-08-03}
}

@article{Weinberg:1958ut,
  title = {Charge {{Symmetry}} of {{Weak Interactions}}},
  author = {Weinberg, Steven},
  date = {1958},
  journaltitle = {Physical Review},
  shortjournal = {Phys. Rev.},
  volume = {112},
  number = {4},
  pages = {1375--1379},
  publisher = {American Physical Society},
  doi = {10.1103/PhysRev.112.1375},
  url = {https://link.aps.org/doi/10.1103/PhysRev.112.1375},
  urldate = {2022-05-16}
}

@article{Schindler:2006jq,
  title = {Nucleon {{Form Factors}} of the {{Isovector Axial-Vector Current}}: {{Situation}} of {{Experiments}} and {{Theory}}},
  shorttitle = {Nucleon {{Form Factors}} of the {{Isovector Axial-Vector Current}}},
  author = {Schindler, M. R. and Scherer, S.},
  date = {2007},
  journaltitle = {The European Physical Journal A},
  shortjournal = {Eur. Phys. J. A},
  volume = {32},
  number = {4},
  eprint = {hep-ph/0608325},
  eprinttype = {arXiv},
  pages = {429--433},
  issn = {1434-6001, 1434-601X},
  doi = {10.1140/epja/i2006-10403-3},
  url = {http://arxiv.org/abs/hep-ph/0608325},
  urldate = {2022-05-16}
}

@article{EuropeanTwistedMass:2008pab,
  title = {Light Baryon Masses with Dynamical Twisted Mass Fermions},
  author = {Alexandrou, C. and Baron, R. and Blossier, B. and Brinet, M. and Carbonell, J. and Dimopoulos, P. and Drach, V. and Farchioni, F. and Frezzotti, R. and Guichon, P. and Herdoiza, G. and Jansen, K. and Korzec, T. and Koutsou, G. and Liu, Z. and Michael, C. and Pène, O. and Shindler, A. and Urbach, C. and Wenger, U.},
  date = {2008},
  journaltitle = {Physical Review D},
  shortjournal = {Phys. Rev. D},
  volume = {78},
  number = {1},
  eprint = {0803.3190},
  eprinttype = {arXiv},
  eprintclass = {hep-lat},
  pages = {014509},
  issn = {1550-7998, 1550-2368},
  doi = {10.1103/PhysRevD.78.014509},
  url = {https://link.aps.org/doi/10.1103/PhysRevD.78.014509},
  urldate = {2025-09-29},
  collaboration = {European Twisted Mass},
  langid = {english}
}

@article{Heo:2022huf,
  title = {The Chiral {{Lagrangian}} with Three Flavors and Large-{{Nc}} Sum Rules},
  author = {Heo, Yonggoo and Kobdaj, C. and Lutz, Matthias F. M.},
  date = {2023-01-03},
  journaltitle = {Eur. Phys. J. A},
  volume = {59},
  number = {1},
  eprint = {2201.04319},
  eprinttype = {arXiv},
  eprintclass = {hep-ph},
  pages = {1},
  doi = {10.1140/epja/s10050-022-00905-5},
  url = {http://arxiv.org/abs/2201.04319},
  urldate = {2022-03-10},
  langid = {english}
}

@article{Leibbrandt:1996np,
  title = {Split {{Dimensional Regularization}} for the {{Coulomb Gauge}}},
  author = {Leibbrandt, George and Williams, Jimmy},
  date = {1996},
  journaltitle = {Nuclear Physics B},
  shortjournal = {Nucl. Phys. B},
  volume = {475},
  number = {1--2},
  eprint = {hep-th/9601046},
  eprinttype = {arXiv},
  pages = {469--483},
  issn = {0550-3213},
  doi = {10.1016/0550-3213(96)00299-4},
  url = {http://arxiv.org/abs/hep-th/9601046},
  urldate = {2025-07-17},
  langid = {english}
}

@article{Haberzettl:1998rw,
    author = "Haberzettl, Helmut",
    title = "{Propagation of a massive spin 3/2 particle}",
    eprint = "nucl-th/9812043",
    archivePrefix = "arXiv",
    month = "12",
    year = "1998",
journal=""
}

@article{Gupta:2024qip,
    author = "Gupta, Rajan",
    title = "{Isovector Axial Charge and Form Factors of Nucleons from Lattice QCD}",
    eprint = "2401.16614",
    archivePrefix = "arXiv",
    primaryClass = "hep-lat",
    reportNumber = "LA-UR-23-341761",
    doi = "10.3390/universe10030135",
    journal = "Universe",
    volume = "10",
    number = "3",
    pages = "135",
    year = "2024"
}

@article{Tsuji:2024scy,
    author = "Tsuji, Ryutaro and Aoki, Yasumichi and Ishikawa, Ken-Ichi and Kuramashi, Yoshinobu and Sasaki, Shoichi and Sato, Kohei and Shintani, Eigo and Watanabe, Hiromasa and Yamazaki, Takeshi",
    title = "{Studies of nucleon isovector structures with the PACS10 superfine lattice}",
    eprint = "2411.16784",
    archivePrefix = "arXiv",
    primaryClass = "hep-lat",
    reportNumber = "KEK-TH-2681",
    doi = "10.22323/1.466.0318",
    journal = "PoS",
    volume = "LATTICE2024",
    pages = "318",
    year = "2025"
}

@article{Passarino:1978jh,
  author = "Passarino, G. and Veltman, M. J. G.",
    title = "{One Loop Corrections for $e^+ e^-$ Annihilation Into $\mu^+ \mu^-$ in the Weinberg Model}",
    reportNumber = "Print-79-0284 (UTRECHT)",
    doi = "10.1016/0550-3213(79)90234-7",
    journal = "Nucl. Phys. B",
    volume = "160",
    pages = "151--207",
    year = "1979"
}

@article{Liang:2022tcj,
    author = "Liang, Ze-Rui and Yao, De-Liang",
    title = "{A unified formulation of one-loop tensor integrals for finite volume effects}",
    eprint = "2207.11750",
    archivePrefix = "arXiv",
    primaryClass = "hep-ph",
    doi = "10.1007/JHEP12(2022)029",
    journal = "JHEP",
    volume = "12",
    pages = "029",
    year = "2022"
}

@article{Lozano:2020qcg,
    author = "Lozano, J. and Agadjanov, A. and Gegelia, J. and Mei\ss{}ner, U. -G. and Rusetsky, A.",
    title = "{Finite volume corrections to forward Compton scattering off the nucleon}",
    eprint = "2010.10917",
    archivePrefix = "arXiv",
    primaryClass = "hep-lat",
    doi = "10.1103/PhysRevD.103.034507",
    journal = "Phys. Rev. D",
    volume = "103",
    number = "3",
    pages = "034507",
    year = "2021",
}

@article{Djukanovic:2022wru,
    author = "Djukanovic, Dalibor and von Hippel, Georg and Koponen, Jonna and Meyer, Harvey B. and Ottnad, Konstantin and Schulz, Tobias and Wittig, Hartmut",
    title = "{Isovector axial form factor of the nucleon from lattice QCD}",
    eprint = "2207.03440",
    archivePrefix = "arXiv",
    primaryClass = "hep-lat",
    reportNumber = "MITP-22-053",
    doi = "10.1103/PhysRevD.106.074503",
    journal = "Phys. Rev. D",
    volume = "106",
    number = "7",
    pages = "074503",
    year = "2022"
}

@article{Alexandrou:2023qbg,
    author = "Alexandrou, Constantia and Bacchio, Simone and Constantinou, Martha and Finkenrath, Jacob and Frezzotti, Roberto and Kostrzewa, Bartosz and Koutsou, Giannis and Spanoudes, Gregoris and Urbach, Carsten",
    collaboration = "Extended Twisted Mass",
    title = "{Nucleon axial and pseudoscalar form factors using twisted-mass fermion ensembles at the physical point}",
    eprint = "2309.05774",
    archivePrefix = "arXiv",
    primaryClass = "hep-lat",
    doi = "10.1103/PhysRevD.109.034503",
    journal = "Phys. Rev. D",
    volume = "109",
    number = "3",
    pages = "034503",
    year = "2024"
}

@article{Ottnad:2022axz,
    author = "Ottnad, Konstantin and Djukanovic, Dalibor and Meyer, Harvey B. and von Hippel, Georg and Wittig, Hartmut",
    title = "{Mass and isovector matrix elements of the nucleon at zero-momentum transfer}",
    eprint = "2212.09940",
    archivePrefix = "arXiv",
    primaryClass = "hep-lat",
    reportNumber = "MITP-22-107",
    doi = "10.22323/1.430.0117",
    journal = "PoS",
    volume = "LATTICE2022",
    pages = "117",
    year = "2023"
}

@article{Harris:2019bih,
    author = "Harris, Tim and von Hippel, Georg and Junnarkar, Parikshit and Meyer, Harvey B. and Ottnad, Konstantin and Wilhelm, Jonas and Wittig, Hartmut and Wrang, Linus",
    title = "{Nucleon isovector charges and twist-2 matrix elements with $N_f=2+1$ dynamical Wilson quarks}",
    eprint = "1905.01291",
    archivePrefix = "arXiv",
    primaryClass = "hep-lat",
    doi = "10.1103/PhysRevD.100.034513",
    journal = "Phys. Rev. D",
    volume = "100",
    number = "3",
    pages = "034513",
    year = "2019"
}

@article{Hall:2025ytt,
    author = "Hall, Zack B. and others",
    title = "{Signs of Non-Monotonic Finite-Volume Corrections to $g_A$}",
    eprint = "2503.09891",
    archivePrefix = "arXiv",
    primaryClass = "hep-lat",
    month = "3",
    year = "2025",
journal=""
}

@article{Park:2025rxi,
    author = "Park, Sungwoo and Gupta, Rajan and Bhattacharya, Tanmoy and He, Fangcheng and Mondal, Santanu and Lin, Huey-Wen and Yoon, Boram",
    title = "{Flavor diagonal nucleon charges using clover fermions on MILC HISQ ensembles}",
    eprint = "2503.07100",
    archivePrefix = "arXiv",
    primaryClass = "hep-lat",
    reportNumber = "LLNL-JRNL-858665, LA-UR-22053",
    month = "3",
    year = "2025",
journal=""
}

@article{Bali:2022qja,
    author = {Bali, Gunnar S. and Collins, Sara and S\"oldner, Wolfgang and Weish\"aupl, Simon},
    collaboration = "RQCD",
    title = "{Leading order mesonic and baryonic SU(3) low energy constants from $N_f$=3 lattice QCD}",
    eprint = "2201.05591",
    archivePrefix = "arXiv",
    primaryClass = "hep-lat",
    doi = "10.1103/PhysRevD.105.054516",
    journal = "Phys. Rev. D",
    volume = "105",
    number = "5",
    pages = "054516",
    year = "2022"
}

@article{Gupta:2018qil,
    author = "Gupta, Rajan and Jang, Yong-Chull and Yoon, Boram and Lin, Huey-Wen and Cirigliano, Vincenzo and Bhattacharya, Tanmoy",
    title = "{Isovector Charges of the Nucleon from 2+1+1-flavor Lattice QCD}",
    eprint = "1806.09006",
    archivePrefix = "arXiv",
    primaryClass = "hep-lat",
    reportNumber = "LA-UR-18-25335, MSUHEP-18-011",
    doi = "10.1103/PhysRevD.98.034503",
    journal = "Phys. Rev. D",
    volume = "98",
    pages = "034503",
    year = "2018"
}

@article{Bali:2023sdi,
    author = {Bali, Gunnar S. and Collins, Sara and Heybrock, Simon and L\"offler, Marius and R\"odl, Rudolf and S\"oldner, Wolfgang and Weish\"aupl, Simon},
    collaboration = "RQCD",
    title = "{Octet baryon isovector charges from $N_f=$2+1 lattice QCD}",
    eprint = "2305.04717",
    archivePrefix = "arXiv",
    primaryClass = "hep-lat",
    doi = "10.1103/PhysRevD.108.034512",
    journal = "Phys. Rev. D",
    volume = "108",
    number = "3",
    pages = "034512",
    year = "2023"
}

@article{Djukanovic:2024krw,
    author = "Djukanovic, Dalibor and von Hippel, Georg and Meyer, Harvey B. and Ottnad, Konstantin and Wittig, Hartmut",
    title = "{Improved analysis of isovector nucleon matrix elements with $N_f$=2+1 flavors of O(a) improved Wilson fermions}",
    eprint = "2402.03024",
    archivePrefix = "arXiv",
    primaryClass = "hep-lat",
    reportNumber = "MITP-24-014",
    doi = "10.1103/PhysRevD.109.074507",
    journal = "Phys. Rev. D",
    volume = "109",
    number = "7",
    pages = "074507",
    year = "2024"
}

@article{Greil:2011aa,
    author = "Greil, Ludwig and Hemmert, Thomas R. and Schafer, Andreas",
    title = "{Finite Volume Corrections to the Electromagnetic Current of the Nucleon}",
    eprint = "1112.2539",
    archivePrefix = "arXiv",
    primaryClass = "hep-ph",
    doi = "10.1140/epja/i2012-12053-2",
    journal = "Eur. Phys. J. A",
    volume = "48",
    pages = "53",
    year = "2012"
}

@book{gradshteyn2014table,
  title     = "Table of integrals, series, and products",
  author    = "Gradshteyn, I S and Ryzhik, I M",
  publisher = "Academic Press",
  month     =  may,
  year      =  2014,
  address   = "San Diego, CA",
  language  = "en"
}

@article{Hasenfratz:1989pk,
  title = {Goldstone boson related finite size effects in field theory and critical phenomena with {O}({N}) symmetry},
	volume = {343},
	issn = {0550-3213},
	url = {https://www.sciencedirect.com/science/article/pii/055032139090603B},
	doi = {10.1016/0550-3213(90)90603-B},
	number = {1},
	journal = {Nuclear Physics B},
	author = {Hasenfratz, P. and Leutwyler, H.},
	month = oct,
	year = {1990},
	pages = {241--284}
}

@article{Hermsen:2024eth,
    author = "Hermsen, Felix and Isken, Tobias and Lutz, Matthias F. M. and Thoma, David",
    title = "{How much strangeness is needed for the axial-vector form factor of the nucleon?}",
    eprint = "2402.04905",
    archivePrefix = "arXiv",
    primaryClass = "hep-ph",
    doi = "10.1103/PhysRevD.109.114029",
    journal = "Phys. Rev. D",
    volume = "109",
    number = "11",
    pages = "114029",
    year = "2024"
}

@article{Isken:2023xfo,
    author = "Isken, Tobias and Guo, Xiao-Yu and Heo, Yonggoo and Korpa, Csaba L. and Lutz, Matthias F. M.",
    title = "{Triangle and box diagrams in coupled-channel systems from the chiral Lagrangian}",
    eprint = "2309.09695",
    archivePrefix = "arXiv",
    primaryClass = "hep-ph",
    doi = "10.1103/PhysRevD.109.034032",
    journal = "Phys. Rev. D",
    volume = "109",
    number = "3",
    pages = "034032",
    year = "2024"
}

@article{Lutz:2023xpi,
    author = "Lutz, Matthias F. M. and Heo, Yonggoo and Guo, Xiao-Yu",
    title = "{Low-energy constants in the chiral Lagrangian with baryon octet and decuplet fields from Lattice QCD data on CLS ensembles}",
    eprint = "2301.06837",
    archivePrefix = "arXiv",
    primaryClass = "hep-lat",
    doi = "10.1140/epjc/s10052-023-11556-1",
    journal = "Eur. Phys. J. C",
    volume = "83",
    number = "5",
    pages = "440",
    year = "2023"
}

@article{Alvarado:2021ibw,
    author = "Alvarado, Fernando and Alvarez-Ruso, Luis",
    title = "{Light-quark mass dependence of the nucleon axial charge and pion-nucleon scattering phenomenology}",
    eprint = "2112.14076",
    archivePrefix = "arXiv",
    primaryClass = "hep-ph",
    doi = "10.1103/PhysRevD.105.074001",
    journal = "Phys. Rev. D",
    volume = "105",
    number = "7",
    pages = "074001",
    year = "2022"
}

@article{Khan:2006de,
author = "Khan, A. Ali and others",
    title = "{Axial coupling constant of the nucleon for two flavours of dynamical quarks in finite and infinite volume}",
    eprint = "hep-lat/0603028",
    archivePrefix = "arXiv",
    reportNumber = "DESY-06-026, EDINBURGH-2006-06, LIVERPOOL-LTH-693, TUM-T39-06-01",
    doi = "10.1103/PhysRevD.74.094508",
    journal = "Phys. Rev. D",
    volume = "74",
    pages = "094508",
    year = "2006"
}

@article{Beane:2004rf,
    author = "Beane, Silas R. and Savage, Martin J.",
    title = "{Baryon axial charge in a finite volume}",
    eprint = "hep-ph/0404131",
    archivePrefix = "arXiv",
    reportNumber = "UNH-04-04, NT-UW-04-08, JLAB-THY-04-14",
    doi = "10.1103/PhysRevD.70.074029",
    journal = "Phys. Rev. D",
    volume = "70",
    pages = "074029",
    year = "2004"
}

@article{Lutz:2001yb,
author = "Lutz, M. F. M. and Kolomeitsev, E. E.",
title = "{Relativistic chiral SU(3) symmetry, large $N_c$ sum rules and meson baryon scattering}",
eprint = "nucl-th/0105042",
archivePrefix = "arXiv",
reportNumber = "GSI-PREPRINT-2001-12, ECT-2001-10",
doi = "10.1016/S0375-9474(01)01312-4",
journal = "Nucl. Phys. A",
volume = "700",
pages = "193--308",
year = "2002"

}

@article{GASSER1988779,
	title = {Nucleons with chiral loops},
	journal = {Nuclear Physics B},
	volume = {307},
	number = {4},
	pages = {779-853},
	year = {1988},
	issn = {0550-3213},
	doi = {https://doi.org/10.1016/0550-3213(88)90108-3},
	url = {https://www.sciencedirect.com/science/article/pii/0550321388901083},
	author = {J. Gasser and M.E. Sainio and A. Švarc}}

@article{Lutz:2020dfi,
    author = "Lutz, Matthias F. M. and Sauerwein, Ulrich and Timmermans, Rob G. E.",
    title = "{On the axial-vector form factor of the nucleon and chiral symmetry}",
    eprint = "2003.10158",
    archivePrefix = "arXiv",
    primaryClass = "hep-lat",
    doi = "10.1140/epjc/s10052-020-8417-5",
    journal = "Eur. Phys. J. C",
    volume = "80",
    number = "9",
    pages = "844",
    year = "2020"
}

@article{Sauerwein:2021jxb,
    author = "Sauerwein, Ulrich and Lutz, Matthias F. M. and Timmermans, Rob G. E.",
    title = "{Axial-vector form factors of the baryon octet and chiral symmetry}",
    eprint = "2105.06755",
    archivePrefix = "arXiv",
    primaryClass = "hep-ph",
    doi = "10.1103/PhysRevD.105.054005",
    journal = "Phys. Rev. D",
    volume = "105",
    number = "5",
    pages = "054005",
    year = "2022"
}

@article{Chen:2012nx,
author = "Chen, Yun-Hua and Yao, De-Liang and Zheng, H. Q.",
title = "{Analyses of pion-nucleon elastic scattering amplitudes up to $O(p^4)$ in extended-on-mass-shell subtraction scheme}",
eprint = "1212.1893",
archivePrefix = "arXiv",
primaryClass = "hep-ph",
doi = "10.1103/PhysRevD.87.054019",
journal = "Phys. Rev. D",
volume = "87",
pages = "054019",
year = "2013"
}

@article{Ando:2006xy,
 author = "Ando, Shung-ichi and Fearing, Harold W.",
title = "{Ordinary muon capture on a proton in manifestly Lorentz invariant baryon chiral perturbation theory}",
eprint = "hep-ph/0608195",
archivePrefix = "arXiv",
reportNumber = "TRI-PP-06-11",
doi = "10.1103/PhysRevD.75.014025",
journal = "Phys. Rev. D",
volume = "75",
pages = "014025",
year = "2007"
}

@article{Ellis:1997kc,
    author = "Ellis, Paul J. and Tang, Hua-Bin",
    title = "{Pion nucleon scattering in a new approach to chiral perturbation theory}",
    eprint = "hep-ph/9709354",
    archivePrefix = "arXiv",
    reportNumber = "NUC-MINN-97-8-T",
    doi = "10.1103/PhysRevC.57.3356",
    journal = "Phys. Rev. C",
    volume = "57",
    pages = "3356--3375",
    year = "1998"
}

@article{Bernard:1993bq, 
author = "Bernard, V. and Kaiser, Norbert and Lee, T. S. H. and Meissner, Ulf-G.",
title = "{Threshold pion electroproduction in chiral perturbation theory}",
eprint = "hep-ph/9310329",
archivePrefix = "arXiv",
reportNumber = "BUTP-93-23, CRN-93-45",
doi = "10.1016/0370-1573(94)90088-4",
journal = "Phys. Rept.",
volume = "246",
pages = "315--363",
year = "1994"
}

@article{Bernard:1998gv,
    author = "Bernard, Veronique and Fearing, Harold W. and Hemmert, Thomas R. and Meissner, Ulf G.",
    title = "{The form-factors of the nucleon at small momentum transfer}",
    eprint = "hep-ph/9801297",
    archivePrefix = "arXiv",
    reportNumber = "KFA-IKP-TH-1998-01, LPT-98-01, TRI-PP-97-73",
    doi = "10.1016/S0375-9474(98)00175-4",
    journal = "Nucl. Phys. A",
    volume = "635",
    pages = "121--145",
    year = "1998",
    note = "[Erratum: Nucl.Phys.A 642, 563--563 (1998)]"
}

@article{Fearing:1997dp,
author = "Fearing, Harold W. and Lewis, Randy and Mobed, Nader and Scherer, Stefan",
title = "{Muon capture by a proton in heavy baryon chiral perturbation theory}",
eprint = "hep-ph/9702394",
archivePrefix = "arXiv",
reportNumber = "TRI-PP-97-5, MKPH-T-97-7",
doi = "10.1103/PhysRevD.56.1783",
journal = "Phys. Rev. D",
volume = "56",
pages = "1783--1791",
year = "1997"
}

@article{Alexandrou:2017hac,
 author = "Alexandrou, Constantia and Constantinou, Martha and Hadjiyiannakou, Kyriakos and Jansen, Karl and Kallidonis, Christos and Koutsou, Giannis and Vaquero Aviles-Casco, Alejandro",
    title = "{Nucleon axial form factors using $N_f$ = 2 twisted mass fermions with a physical value of the pion mass}",
    eprint = "1705.03399",
    archivePrefix = "arXiv",
    primaryClass = "hep-lat",
    reportNumber = "DESY-17-064",
    doi = "10.1103/PhysRevD.96.054507",
    journal = "Phys. Rev. D",
    volume = "96",
    number = "5",
    pages = "054507",
    year = "2017"
}

@article{Lutz:2014oxa, 
author = "Lutz, M. F. M. and Bavontaweepanya, R. and Kobdaj, C. and Schwarz, K.",
title = "{Finite volume effects in the chiral extrapolation of baryon masses}",
eprint = "1401.7805",
archivePrefix = "arXiv",
primaryClass = "hep-lat",
doi = "10.1103/PhysRevD.90.054505",
journal = "Phys. Rev. D",
volume = "90",
number = "5",
pages = "054505",
year = "2014"
}

@article{Bali:2014nma,
  author = {Bali, Gunnar S. and Collins, Sara and Gl\"assle, Benjamin and G\"ockeler, Meinulf and Najjar, Johannes and R\"odl, Rudolf H. and Sch\"afer, Andreas and Schiel, Rainer W. and S\"oldner, Wolfgang and Sternbeck, Andr\'e},
    title = "{Nucleon isovector couplings from $N_f=2$ lattice QCD}",
    eprint = "1412.7336",
    archivePrefix = "arXiv",
    primaryClass = "hep-lat",
    doi = "10.1103/PhysRevD.91.054501",
    journal = "Phys. Rev. D",
    volume = "91",
    number = "5",
    pages = "054501",
    year = "2015",
}

@article{Semke:2005sn,
author = "Semke, A. and Lutz, M. F. M.",
title = "{Baryon self energies in the chiral loop expansion}",
eprint = "nucl-th/0511061",
archivePrefix = "arXiv",
doi = "10.1016/j.nuclphysa.2006.07.043",
journal = "Nucl. Phys. A",
volume = "778",
pages = "153--180",
year = "2006"
}

@article{Alexandrou:2008tn,
author = "Alexandrou, C. and others",
collaboration = "European Twisted Mass",
title = "{Light baryon masses with dynamical twisted mass fermions}",
eprint = "0803.3190",
archivePrefix = "arXiv",
primaryClass = "hep-lat",
reportNumber = "LPT-ORSAY-08-32, IRFU-08-29, DESY-08-032, SFB-CPP-08-19, ROM2F-2008-06",
doi = "10.1103/PhysRevD.78.014509",
journal = "Phys. Rev. D",
volume = "78",
pages = "014509",
year = "2008"
}

@article{Bali:2018qus,
author = {Bali, G. S. and Collins, S. and Gruber, M. and Sch\"afer, A. and Wein, P. and Wurm, T.},
title = "{Solving the PCAC puzzle for nucleon axial and pseudoscalar form factors}",
eprint = "1810.05569",
archivePrefix = "arXiv",
primaryClass = "hep-lat",
doi = "10.1016/j.physletb.2018.12.053",
journal = "Phys. Lett. B",
volume = "789",
pages = "666--674",
year = "2019"
}

@article{Capitani:2015sba,
 author = {Capitani, S. and Della Morte, M. and Djukanovic, D. and von Hippel, G. and Hua, J. and J\"ager, B. and Knippschild, B. and Meyer, H. B. and Rae, T. D. and Wittig, H.},
title = "{Nucleon electromagnetic form factors in two-flavor QCD}",
eprint = "1504.04628",
archivePrefix = "arXiv",
primaryClass = "hep-lat",
reportNumber = "MITP-15-026, HIM-2015-01, CP3-ORIGINS-2015-012, DIAS-2015-12",
doi = "10.1103/PhysRevD.92.054511",
journal = "Phys. Rev. D",
volume = "92",
number = "5",
pages = "054511",
year = "2015"
}

@article{Procura:2006gq,
    author = "Procura, M. and Musch, B. U. and Hemmert, T. R. and Weise, W.",
    title = "{Chiral extrapolation of $g_A$ with explicit $\Delta$(1232) degrees of freedom}",
    eprint = "hep-lat/0610105",
    archivePrefix = "arXiv",
    reportNumber = "TUM-T39-06-10",
    doi = "10.1103/PhysRevD.75.014503",
    journal = "Phys. Rev. D",
    volume = "75",
    pages = "014503",
    year = "2007"
}

@article{Jenkins:1990jv,
 author = "Jenkins, Elizabeth Ellen and Manohar, Aneesh V.",
title = "{Baryon chiral perturbation theory using a heavy fermion Lagrangian}",
reportNumber = "UCSD-PTH-90-23",
doi = "10.1016/0370-2693(91)90266-S",
journal = "Phys. Lett. B",
volume = "255",
pages = "558--562",
year = "1991"
}

@article{Fuchs:2003vw, 
author = "Fuchs, T. and Scherer, S.",
title = "{Pion electroproduction, PCAC, chiral ward identities, and the axial form-factor revisited}",
eprint = "nucl-th/0303002",
archivePrefix = "arXiv",
reportNumber = "MKPH-T-03-4",
doi = "10.1103/PhysRevC.68.055501",
journal = "Phys. Rev. C",
volume = "68",
pages = "055501",
year = "2003"
}

@article{Ledwig:2011cx,
author = "Ledwig, T. and Martin-Camalich, J. and Pascalutsa, V. and Vanderhaeghen, M.",
title = "{The Nucleon and $\Delta$(1232) form factors at low momentum-transfer and small pion masses}",
eprint = "1108.2523",
archivePrefix = "arXiv",
primaryClass = "hep-ph",
reportNumber = "MKPH-T-11-15",
doi = "10.1103/PhysRevD.85.034013",
journal = "Phys. Rev. D",
volume = "85",
pages = "034013",
year = "2012"
}

@article{Jenkins:1991es,
author = "Jenkins, Elizabeth Ellen and Manohar, Aneesh V.",
title = "{Chiral corrections to the baryon axial currents}",
reportNumber = "UCSD-PTH-91-05",
doi = "10.1016/0370-2693(91)90840-M",
journal = "Phys. Lett. B",
volume = "259",
pages = "353--358",
year = "1991"
}

@article{Alexandrou:2010hf,
author = "Alexandrou, C. and Brinet, M. and Carbonell, J. and Constantinou, M. and Harraud, P. A. and Guichon, P. and Jansen, K. and Korzec, T. and Papinutto, M.",
collaboration = "ETM",
title = "{Axial Nucleon form factors from lattice QCD}",
eprint = "1012.0857",
archivePrefix = "arXiv",
primaryClass = "hep-lat",
doi = "10.1103/PhysRevD.83.045010",
journal = "Phys. Rev. D",
volume = "83",
pages = "045010",
year = "2011"
}

@article{Yao:2017fym,   author = "Yao, De-Liang and Alvarez-Ruso, Luis and Vicente-Vacas, Manuel J.",
title = "{Extraction of nucleon axial charge and radius from lattice QCD results using baryon chiral perturbation theory}",
eprint = "1708.08776",
archivePrefix = "arXiv",
primaryClass = "hep-ph",
doi = "10.1103/PhysRevD.96.116022",
journal = "Phys. Rev. D",
volume = "96",
number = "11",
pages = "116022",
year = "2017"
}

@article{Schindler:2006it,  author = "Schindler, M. R. and Fuchs, T. and Gegelia, J. and Scherer, S.",
title = "{Axial, induced pseudoscalar, and pion-nucleon form-factors in manifestly Lorentz-invariant chiral perturbation theory}",
eprint = "nucl-th/0611083",
archivePrefix = "arXiv",
reportNumber = "MKPH-T-06-18",
doi = "10.1103/PhysRevC.75.025202",
journal = "Phys. Rev. C",
volume = "75",
pages = "025202",
year = "2007"
}

@article{Fuchs:2003qc,
author = "Fuchs, T. and Gegelia, J. and Japaridze, G. and Scherer, S.",
title = "{Renormalization of relativistic baryon chiral perturbation theory and power counting}",
eprint = "hep-ph/0302117",
archivePrefix = "arXiv",
reportNumber = "MKPH-T-03-2",
doi = "10.1103/PhysRevD.68.056005",
journal = "Phys. Rev. D",
volume = "68",
pages = "056005",
year = "2003"
}

@article{Lutz:2018cqo,
 author = "Lutz, M. F. M. and Heo, Yonggoo and Guo, Xiao-Yu",
title = "{On the convergence of the chiral expansion for the baryon ground-state masses}",
eprint = "1801.06417",
archivePrefix = "arXiv",
primaryClass = "hep-lat",
doi = "10.1016/j.nuclphysa.2018.05.007",
journal = "Nucl. Phys. A",
volume = "977",
pages = "146--207",
year = "2018"
}

@article{Capitani:2017qpc,
 author = {Capitani, Stefano and Della Morte, Michele and Djukanovic, Dalibor and von Hippel, Georg M. and Hua, Jiayu and J\"ager, Benjamin and Junnarkar, Parikshit M. and Meyer, Harvey B. and Rae, Thomas D. and Wittig, Hartmut},
    title = "{Isovector axial form factors of the nucleon in two-flavor lattice QCD}",
    eprint = "1705.06186",
    archivePrefix = "arXiv",
    primaryClass = "hep-lat",
    reportNumber = "CP3-Origins-2017-018, HIM-2017-03, MITP/17-029, TIFR/TH/17-21, CP3-ORIGINS-2017-018, MITP-17-029, TIFR-TH-17-21",
    doi = "10.1142/S0217751X1950009X",
    journal = "Int. J. Mod. Phys. A",
    volume = "34",
    number = "02",
    pages = "1950009",
    year = "2019"
}

@article{Hemmert:2003cb,
    author = "Hemmert, Thomas R. and Procura, Massimiliano and Weise, Wolfram",
    title = "{Quark mass dependence of the nucleon axial vector coupling constant}",
    eprint = "hep-lat/0303002",
    archivePrefix = "arXiv",
    reportNumber = "TUM-T39-03-03",
    doi = "10.1103/PhysRevD.68.075009",
    journal = "Phys. Rev. D",
    volume = "68",
    pages = "075009",
    year = "2003"
}

\end{document}